\documentclass{aa}     % LaTeX A&A New Fonts
\usepackage{epsfig}

\def\a4{Abell 496}

\thesaurus{11.03.4 Abell 496; 11.03.1 }

\title{Optical and X-ray analysis of the cluster of galaxies Abell~496.
\thanks{Based on observations collected at the European Southern
Observatory, La Silla, Chile}
}

 \author {
  F.~Durret \inst{1,2}
\and
  C.~Adami \inst{3,4}
\and
  D.~Gerbal \inst{1,2}
\and
  V.~Pislar  \inst{1,5}
}
\offprints{F.~Durret, durret@iap.fr }
\institute{
  Institut d'Astrophysique de Paris, CNRS, 
  98bis Bd Arago, F-75014 Paris, France 
\and 
    DAEC, Observatoire de Paris, Universit\'e Paris VII, CNRS (UA 173),
    F-92195 Meudon Cedex, France 
\and
IGRAP, Laboratoire d'Astronomie Spatiale, 
Traverse du Siphon, F-13012 Marseille, France  
\and 
Department of Physics and Astronomy, Northwestern University, Dearborn
Observatory, 2131 Sheridan, 60208-2900 Evanston, USA
\and
Universit\'e du Havre, 25 rue Philippe Lebon, 76600 Le Havre, France
}
\date{Received 17 December 1999/ Accepted}
\begin{document}

\maketitle

\begin{abstract}

We present a detailed analysis of the cluster of galaxies \a4.  The
optical data include a redshift catalogue of 466 galaxies, out of
which 274 belong to the main cluster and a CCD photometric catalogue
in a much smaller region, with 239 and 610 galaxies in the V and R
bands respectively. The X-ray analysis is based on an image obtained
with the ROSAT PSPC. \\ Besides Abell 496 itself, the velocity
distribution along the line of sight shows the existence of at least
four structures at different redshifts, one of them seeming to be a
poor cluster at a velocity of 30083 km/s. The other of these
structures have a too large spatial extent to be clusters but may be
filaments along the line of sight or other young structures.\\ Various
independent methods show that Abell 496 appears to be a quite relaxed
cluster, except perhaps for the distribution of emission line
galaxies.  These appear to be distributed in two samples falling on to
the main cluster, one from the back (the ELGs concentrated towards the
west) and one from the front (the high velocity ELGs). \\ The bright
part of the galaxy luminosity function, built from the redshift
catalogue, shows a flattening at R$\sim 16$ (M$_{\rm R}\sim -20.5$),
and can be accounted for by a gaussian distribution of bright galaxies
and a power law or Schechter function for faint galaxies. The deeper
galaxy counts derived from CCD imaging show a dip at R$\sim 19.5$
(M$_{\rm R} \sim -17$) which can be modelled assuming a cut-off in the
luminosity function such as that observed in Coma.\\ We propose a
model for the X-ray gas and derive the galaxy, X-ray gas and total
dynamical masses, as well as the baryon fraction in the cluster. \a4\
appears as a relaxed cluster which can be used as a prototype for
further studies.\\

\end{abstract}

\section{Introduction}

In the framework of hierarchical clustering, the Universe is believed
to be made of galaxies distributed in sheets encircling voids or
filaments, at the intersection of which clusters of galaxies are
located. Such models can be tested through the analysis of clusters,
which are likely to keep a ``memory'' of their formation. This is
suggested for example by the alignment effects observed in some
clusters, such as for example Abell 3558 (Dantas et al. 1997) or Abell
85 (Durret et al. 1998), where the cD, the brightest galaxies, the
X-ray emitting gas and possibly even larger scale structures (in the
case of Abell 85) all appear aligned along the same direction.
Multi-wavelength studies of clusters of galaxies also allow us to draw
a global and coherent portrait of these objects, which we can then use
to address other questions of interest, such as the influence of
mergers and environmental effects at various scales on the properties
of both galaxies and X-ray gas. Large scale (i.e. cluster size)
mergers are quite often observed from substructures detected in the
X-ray gas; smaller scale mergers (i.e. group size) such as the infall
of dwarf galaxies onto groups surrounding bright galaxies can be
derived from various methods such as those of Serna \& Gerbal (1996)
or Gurzadyan \& Mazure (1998), which require optical velocity and
magnitude catalogues; the existence of subclustering also has an
influence on the shape of the galaxy luminosity function, which in
some cases appears to show a deficit of faint galaxies often
interpreted as due to accretion of dwarf galaxies onto larger galaxies
or groups (e.g. in Coma, Lobo et al. 1997, Adami et al. 2000).  It
therefore appears important to analyze cluster properties in detail
before using them in other studies.  Note in particular that the
existence of substructures can lead to overestimate cluster velocity
dispersions, and hence M/L ratios and the value of $\Omega _0$ in
clusters.

With the improvement of both observational means (better X-ray
detectors, optical multi-object spectroscopy) and modern methods of
analysis (some of which are described below), an ever increasing
number of clusters showing evidence for merging and environmental
effects has been found. A rather general picture has therefore emerged
for clusters, with a main relaxed body on to which groups of various
sizes can be falling.

Our approach these last years has been to study a small sample of
nearby clusters in detail. These have the advantage of being bright,
and can therefore be observed in detail within a reasonable amount of
telescope time. Besides, they are free of evolution effects.  We
present here a detailed multi-wavelength study of \a4, based on
optical (extensive redshift and photometric catalogues) and X-ray
(ROSAT PSPC) data.

\a4 is a richness class 1 (Abell 1958) cD type (Struble \& Rood 1987)
cluster at a redshift of 0.0331. For a Hubble constant H$_0$=50 km
s$^{-1}$ Mpc$^{-1}$, the corresponding scale is 55.0 kpc/arcmin and
the distance modulus is 36.52. At optical wavelengths, an adaptive
kernel map of the central region (in a 60$\times$60 arcmin$^2$ square)
has revealed a somewhat complicated structure, with a strong
concentration of galaxies in the north-south direction (Kriessler \&
Beers 1997). Note however that this map does not include redshift
information. Thorough investigations of the X-ray properties of \a4\
can be found in Mohr et al. (1999) and Markevitch et al. (1999); their
results will be compared to ours in section 4.5.

The paper is organized as follows: we present the data in section 2;
the structures along the line of sight derived from the velocity
catalogue are described in section 3; the optical properties of the
\a4\ cluster itself are described in section 4; the X-ray cluster
properties are described in section 5; a summary and conclusions are
given in section 6.

\section{The data} 

\subsection{Optical data}

Our redshift catalogue includes 466 galaxies in the direction of the
cluster \a4, in a region covering about 160$\times$160 arcmin$^2$
(9.2$\times$9.2~Mpc for an average redshift for \a4 of 0.0331). It is
fully described in Durret et al. (1999b).

Our photometric catalogues are described in Slezak et al. (1999).  The
photographic plate catalogue was obtained by scanning an SRC plate in
the b$_{\rm J}$ band with the MAMA measuring machine at the
Observatoire de Paris; it includes 3879 galaxies located in a region
of roughly $\pm$~1.3$^\circ$ from the cluster centre.  Positions are
very accurate in this catalogue and were used for spectroscopy; on the
other hand, b$_{\rm J}$ magnitudes are not accurate, so a CCD
photometric catalogue was obtained in the V and R bands in order to
recalibrate photographic plate magnitudes. The R magnitudes thus
estimated for the photographic plate catalogue were used to estimate
the completeness of our redshift catalogue. The CCD imaging catalogue
includes 239 and 610 galaxies in the V and R bands respectively, and
is limited to a much smaller region of $\sim 246$ arcmin$^2$ in the
centre of the cluster.

The cluster center will be taken to be the position of the cD galaxy,
which coincides with the X-ray maximum: $\alpha = 4^h 33^{mn} 38.45^s,
\delta = -13^\circ 15' 49.5''$ (2000.0). 

\subsection{X-ray data}

The ROSAT PSPC image (\#800024) was taken from the archive and
analyzed by Pislar (1998). The cluster was observed during 8972~s.
The effective exposure time for this image, after data reduction is
5354~s. We have used the routines developed by Snowden (1995) to
obtain a non cosmic background subtracted image between 0.5 and
2~keV.  We have defined the image limiting radius
($\sim\,1620\,h_{50}^{-1}$ kpc) as the radius where the surface
brightness reaches the surface brightness of the cosmic background
($3\,10^{-4}\,{\rm{s}}^{-1}\,{\rm{arcmin}}^{-2}$). The different PSPC
background components are detailed in Snowden et al. (1994).

The global X-ray gas temperatures derived from Einstein and EXOSAT
satellite data are 3.9$\pm$0.2 keV (David et al. 1993) and
$4.7^{+1}_{-0.8}$~keV respectively (Edge \& Stewart 1991).

\section{Velocity distribution along the line of sight}

We first discuss the overall properties derived from the velocity
distribution along the line of sight.

\subsection{Overall characteristics of the structures detected along
the line of sight}

\begin{figure}[h]
\centerline{\psfig{figure=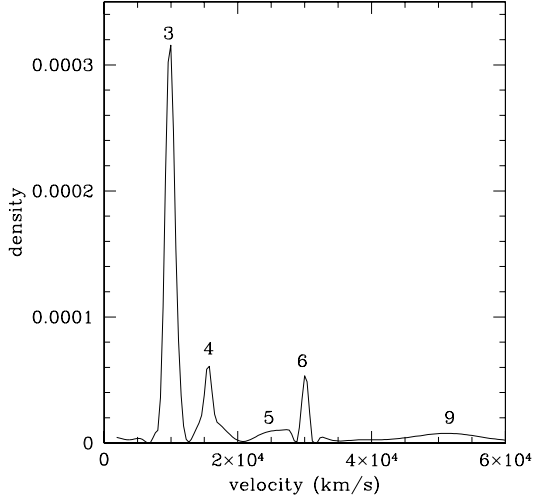,height=7cm}}
\caption[ ]{Wavelet reconstruction of the velocity distribution in the
direction of \a4. The numbers above the peaks correspond to those of the
structures described in the text and in Table \ref{tab9groupes}. The
density units correspond to a total integrated galaxy density of 1.}
\protect\label{pdfall}
\end{figure}

A wavelet reconstruction of the velocity distribution along the line
of sight is displayed in Fig. \ref{pdfall} (466 galaxies). We remind
the reader that this type of reconstruction takes into account
structures at a significance level of at least 3$\sigma$, and detects
structures at various scales.  The sample was analyzed with 256
points, and the smallest scale was excluded because it is very noisy.
A more detailed description of this technique can be found in Fadda et
al. (1998).

Nine ``groups'' or velocity substructures are found with this method,
and their velocity characteristics are given in Table
\ref{tab9groupes}.  The group number is given in col. 1, the number of
galaxies in col. 2, the BWT mean velocity (Beers et al. 1990) and
corresponding BWT velocity dispersion in cols. 3 and 4, and the
velocity interval in col. 5. Foreground groups 1 and 2 and background
groups 7 and 8 are most probably not real groups, since they are
widely spread both on the sky and in velocity distribution; because of
the small number of galaxies involved in the first three of these
groups, we did not calculate mean velocities or velocity dispersions
for these structures. For group 8 these values are only indicative,
but characteristic of a low mass structure.  Group 3 appears to be the
cluster \a4\ itself. Except for 7 objects, all the galaxies in group 4
appear to be located north of \a4. Group 5 has the same kind of shape
and size as group 4 and is roughly coaligned with \a4\ along the line
of sight.  Group 6 appears strongly concentrated both spatially and in
velocity space, all but two galaxies being located west of
\a4. Moreover, its velocity dispersion is also low.

\begin{table}[h]
\caption{Structures detected along the line of sight}
\begin{tabular}{rrrrc}
\hline
Name & N$_{gal}$ & v$_{mean}$ & $\sigma _v$~ & velocity interval \\
     &           & (km/s)~    & (km/s)       & (km/s)~~~ \\
\hline
1 &   6 &       &      & [1803,3396] \\
2 &   6 &       &      & [5124,6233] \\
3 & 274 &  9885 &  715 & [7813,11860] \\
4 &  59 & 15758 & 1214 & [13458,17928] \\
5 &  25 & 25361 & 1627 & [23477,28209] \\
6 &  29 & 30083 &  380 & [29144,30814] \\
7 &   5 &       &      & [40606,41650] \\
8 &   9 & 46349 &  263 & [45836,46660] \\
9 &  22 & 52358 & 1467 & [50052,54727] \\
\hline
\end{tabular}
\protect\label{tab9groupes}
\end{table}

These results are confirmed when we apply the same method as for the
ENACS clusters (Katgert et al. 1996, Mazure et al. 1996) to detect
velocity structures along the line of sight. This method consists in
sorting the galaxies in order of increasing velocity, and plotting
their rank as a function of velocity (hereafter the rank-velocity
classification). If the distribution of galaxies in redshift space is
strictly gaussian, we expect to see a regular S-shape in the
sequence/gap space. When there are more than 5 galaxies between two
successive gaps, we consider that the galaxies belong to a structure.

\subsection{A finer analysis of structures 4, 5, 6 and 9}

\begin{figure}[h!]
\centerline{\psfig{figure=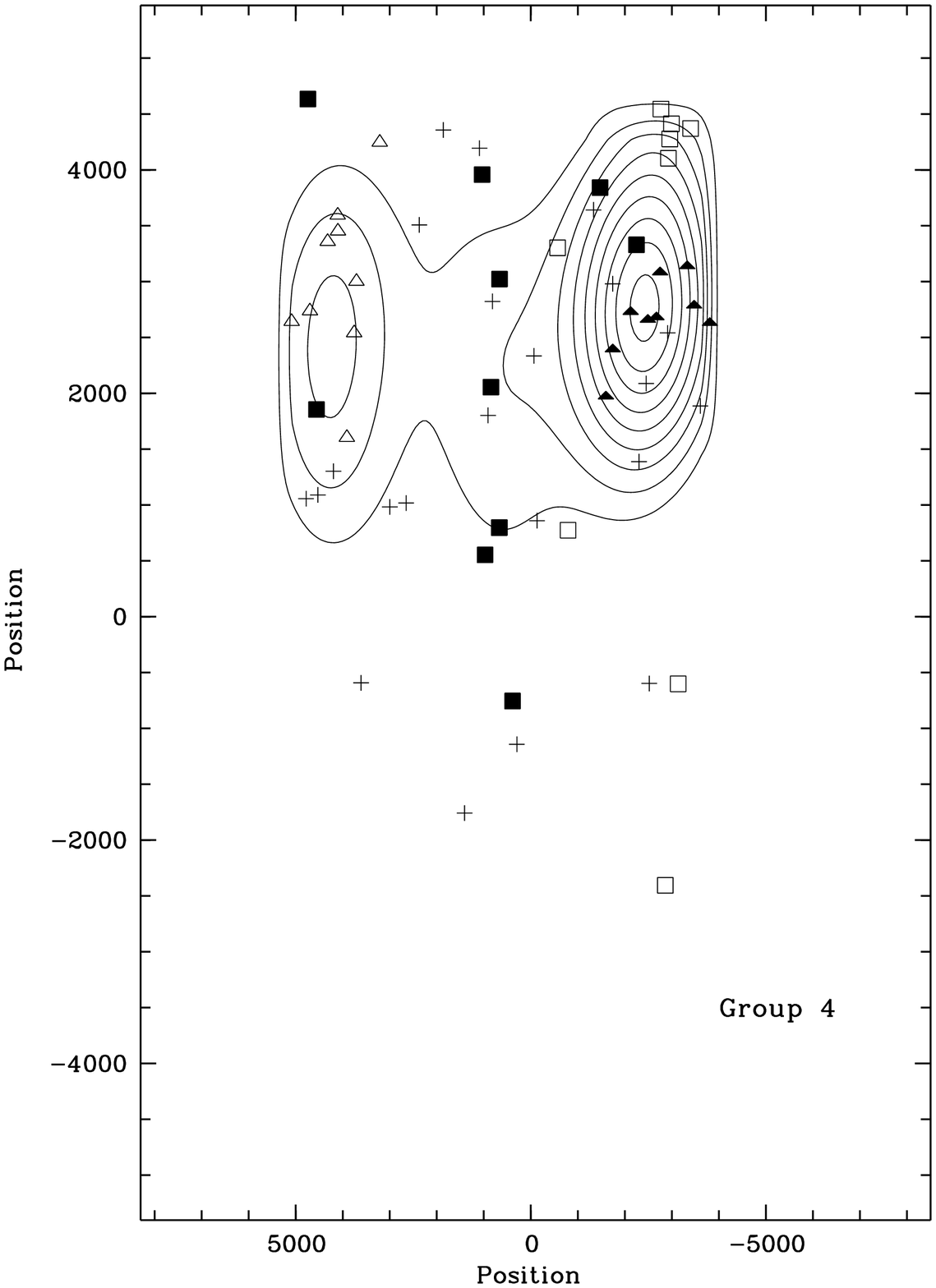,width=8.1cm,height=4.8cm,clip=true}}
\centerline{\psfig{figure=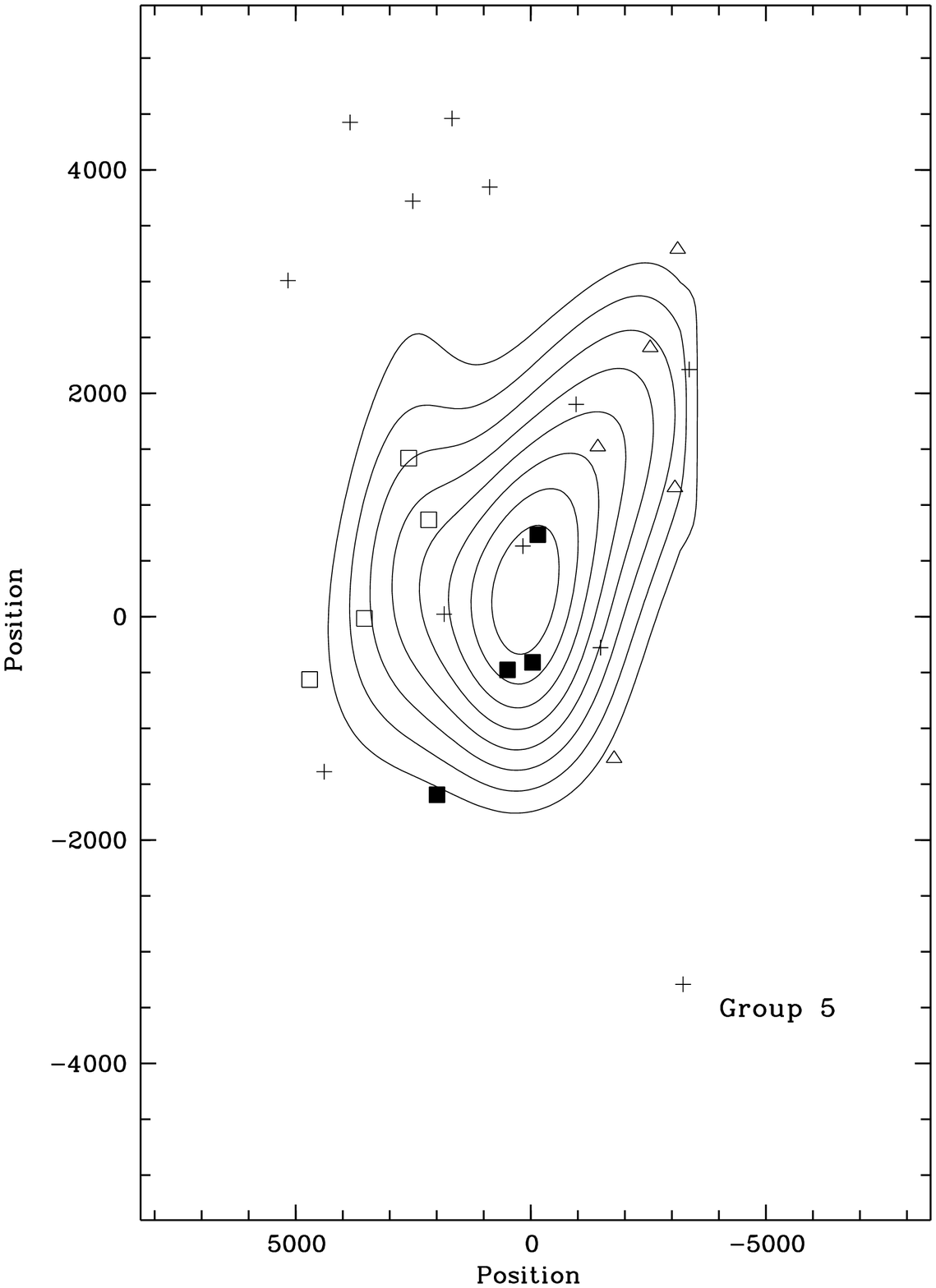,width=8.1cm,height=4.8cm,clip=true}}
\centerline{\psfig{figure=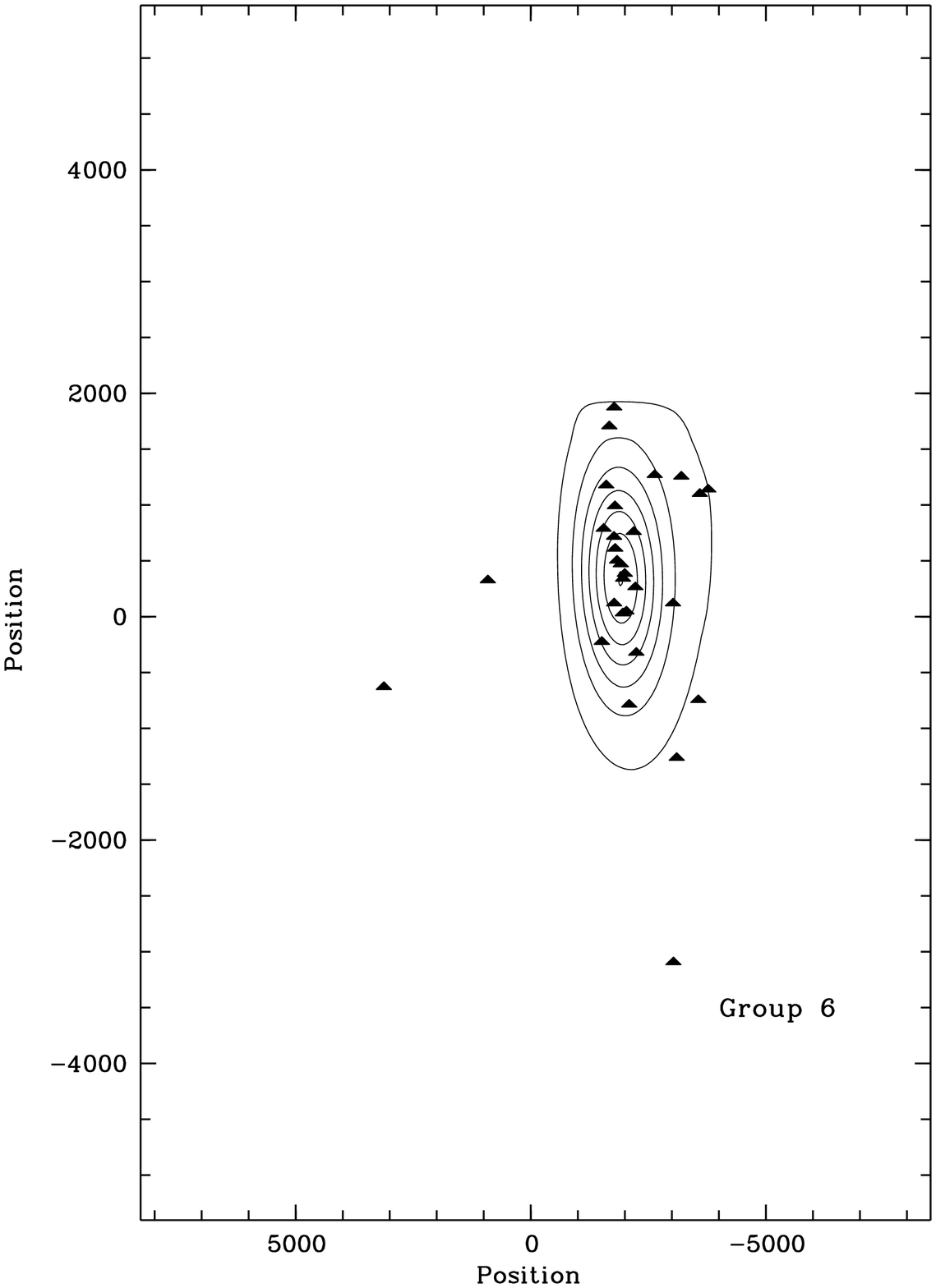,width=8.1cm,height=4.8cm,clip=true}}
\centerline{\psfig{figure=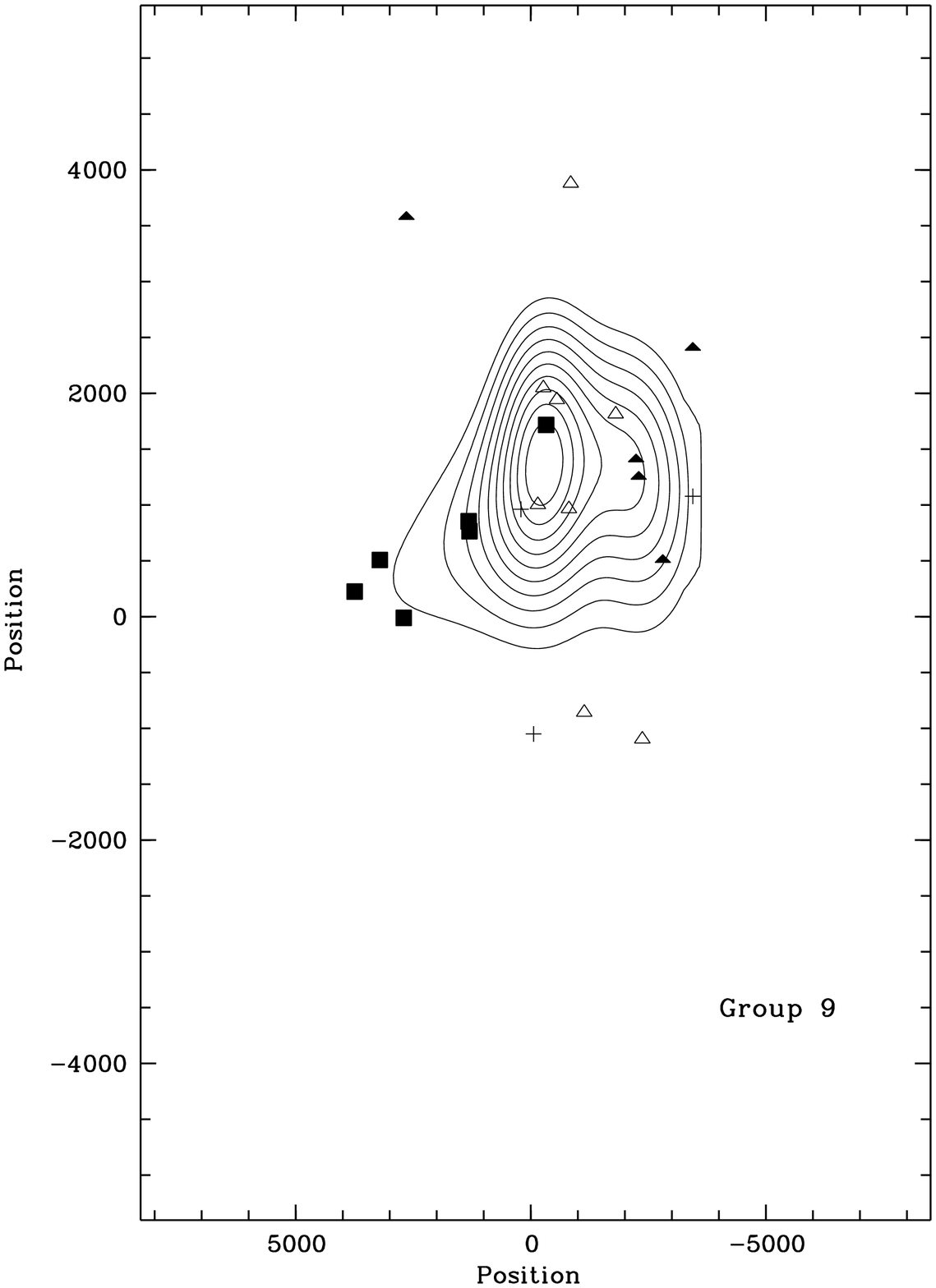,width=8.1cm,height=4.8cm,clip=true}}
\caption[ ]{Isodensity contours for galaxies in structures 4, 5, 6 and
9 (from top to bottom); galaxy positions are superimposed with the
following symbols: structure 4: empty rectangles=4a, filled
rectangles=4b, empty triangles=4c, filled triangles=4d, crosses=other
galaxies in structure 4; structure 5: empty rectangles=5a, filled
rectangles=5b, empty triangles=5c, crosses=other galaxies in structure
5; structure 6: all galaxies; structure 9: filled rectangles=9a,
filled triangles=9b, empty triangles=9c, crosses=other galaxies in
structure~9. Positions are relative to the cluster center defined in the text.}
\protect\label{xyg4569}
\end{figure}

In order to understand whether groups 4, 5, 6 and 9 can be physically
coherent structures, we performed a Serna \& Gerbal (1996) analysis
for each of these groups separately. Since this type of analysis takes
into account galaxy magnitudes, we had to eliminate one galaxy in
group 4 and one in group 5, for which we have no magnitude in our
redshift catalogue. We also tried keeping these galaxies and assigning
them an ``average'' magnitude R=17. The results in both cases were
similar. Note that the redshift catalogue completeness is about
50\% in regions 4 and 5, and 55\% in region 9, within the magnitude
limit R=18.8. It could therefore be argued that the Serna \& Gerbal
method is meaningless for these samples. However, in this type of
analysis it is the brightest galaxies which mainly contribute to the
dynamics of the system (since the mass to luminosity ratio is taken to
be constant for all galaxies). If the samples are limited to 
magnitudes R$\leq$17.0, the redshift catalogue completeness then
becomes 72\%, 72\% and 79\% for regions 4, 5 and 9 respectively, and
the Serna \& Gerbal method is therefore fully applicable.

\begin{table}[h]
\caption{Substructures detected along the line of sight}
\begin{tabular}{rrrr}
\hline
Name & N$_{gal}$ & v$_{mean}$ & $\sigma _v$~ \\
     &           & (km/s)~    & (km/s)       \\
\hline
4a &  9 & 14043 & 381 \\
4b & 10 & 17568 & 243 \\
4c &  9 & 15302 & 201 \\
4d &  9 & 15604 & 131 \\
5a &  4 & 23785 & 195 \\ 
5b &  4 & 25159 & 174 \\
5c &  5 & 26719 & 457 \\
9a &  6 & 54164 & 119 \\
9b &  5 & 51165 &  28 \\
9c &  8 & 52013 & 348 \\
\hline
\end{tabular}
\protect\label{tabssgroupes}
\end{table}

The characteristics of the substructures found with the Serna \&
Gerbal method are given in Table \ref{tabssgroupes}.  Structure 4
has subgroups 4c and 4d well defined in space; they extend over 5
and 3.6 Mpc respectively and could therefore be members of two
different clusters. Due to their large spatial extent, subgroups 4a
and 4b respectively seem to be just forward and background galaxies,
with the exception of the five galaxies of group 4a at the north west
extremity (see Fig. \ref{xyg4569}).

Subgroups 5a and 5b form structures with a small velocity dispersion,
but extending over about 7 Mpc, a size which appears rather large for
these groups to be members of two background clusters; the extent and
the velocity dispersion of 5c are even larger (Fig. \ref{xyg4569}).

While the Serna \& Gerbal (1996) method finds dynamical sub-structures
for the other groups along the line of sight, the same method reveals
no substructures in group 6, except for two pairs of galaxies; group 6
therefore appears well defined both in velocity distribution and in
space. Eighteen galaxies are included in an ellipse with a major and
minor axes of about 8 and 3 Mpc, suggesting that this is a poor,
diffuse and low mass cluster (Fig. \ref{xyg4569}).

\begin{figure}[h!]
\centerline{\psfig{figure=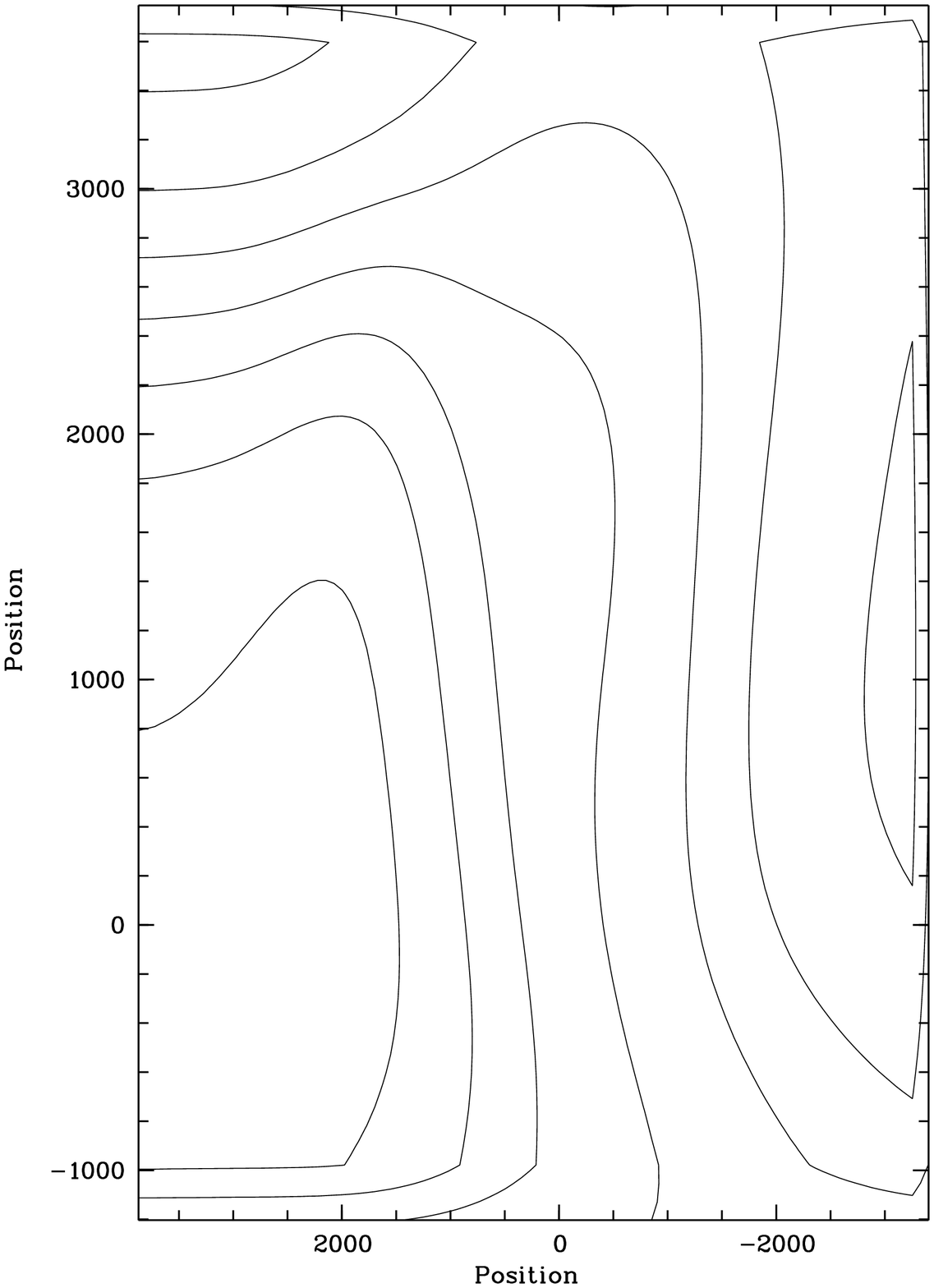,height=5cm,width=7.9cm}}
\caption[ ]{Isocontours of the velocity field in group 9, from 51000 
(right) to 54000 km/s (left) by steps of 500 km/s. Positions are relative 
to the cluster center defined in the text.}
\protect\label{vg9}
\end{figure}

Finally, three subgroups are apparently found in structure 9, 9a and
9b having very small velocity dispersions but spanning a rather large
spatial region. The overall velocity field in group 9 shows an
interesting pattern looking like a velocity gradient
(Fig. \ref{vg9}). This could be a filament more or less aligned along
the line of sight.

A rank-velocity classification applied to each group confirms that
group 4 and possibly group 9 appear to have three substructures (two
clear breaks in the curves), group 5 has three or four substructures
and group 6 has no clear substructure except perhaps for the two or
three first galaxies which are probably infalling objects.

To summarize, groups 4 and 9 clearly appear as filaments or at least
elongated structures along the line of sight, but not really massive
clusters. This analysis is confirmed by the iso-velocity contours of
group 9. The continuous velocity gradient could be interpreted as the
result of a merger, with the infalling groups not perfectly aligned
along the line of sight. Group 6 has a low velocity dispersion and is
probably a poor cluster. Group 5 exhibits two low velocity dispersion
sub-structures and a moderately high velocity dispersion group
(5c), but the number of galaxies in 5c is too low to provide a robust
estimation and we assume that this structure is not a cluster.

We will now discuss the dynamical state of the main structure on the
line of sight: the cluster \a4\ itself (group 3).

\section{Morphological and physical properties of Abell 496}

\subsection{Morphology of the cluster at various wavelengths}

We display in Fig. \ref{image} the superposition of the optical image
of the cluster with ROSAT PSPC X-ray and radio isocontours. The X-ray
contours are quite smooth, with no obvious substructures. However,
there appears to be an excess of X-ray emission towards the north
west, in the direction where there is also an excess of emission line
galaxies (see below). A radio source is visible south east of the
cluster, probably associated with a galaxy.

\begin{figure}[h!]
\centerline{\psfig{figure=fig4.ps,height=7cm}}
\caption[ ]{Optical POSS image of the central region of Abell 496 on to 
which X-ray (full lines) and radio 1400 MHz (dot-dashed lines) contours are 
superimposed. The radio data were taken from Condon et al. (1998). }
\protect\label{image}
\end{figure}

\subsection{The galaxy velocity distribution in Abell 496}

\begin{figure}[b]
\centerline{\psfig{figure=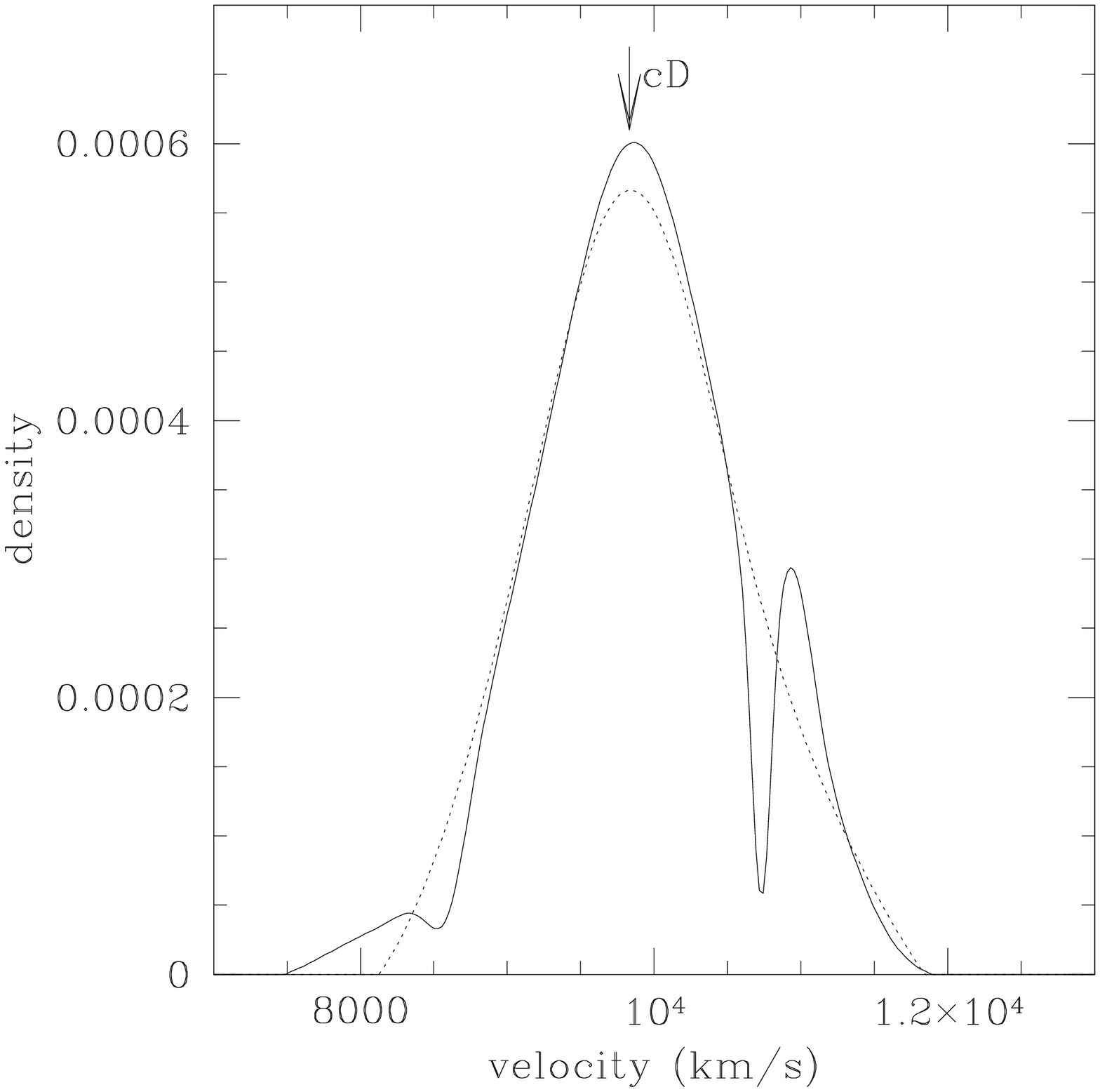,height=7cm}}
\caption[ ]{Wavelet reconstruction of the velocity distribution in the
direction of \a4, obtained by excluding the two smallest scales (full line)
and the three smallest scales (dashed line). The arrow indicates the 
velocity of the cD galaxy. The
density units correspond to a total integrated galaxy density of 1.}
\protect\label{pdfin}
\end{figure}

The cluster \a4\ corresponds to structure 3 in Table
\ref{tab9groupes}; it has a BWT mean velocity of 9885 km/s and a
global velocity dispersion of 715 km/s. The corresponding velocity
interval is [7813,11860 km/s] and includes 274 galaxies.  Note that
the cD galaxy has a velocity of 9831 km/s, close to the mean cluster
velocity, and is located very close the X-ray emission center,
suggesting that the cD is at the bottom of the cluster gravitational
potential well.  This is an indication of a quiescent history of the
cluster (see e.g. Zabludoff et al. 1993, Oegerle \& Hill 1994).

The wavelet reconstruction of the velocity distribution of \a4\ shown
in Fig. \ref{pdfin} (274 galaxies) suggests the presence of a certain
amount of substructure. The sample was analyzed with 256 points, and
the two smallest scales were excluded.  The corresponding velocity
distribution is non-gaussian; it shows: a tiny feature at $\sim$8300
km/s; a main asymmetric peak in the [8500, 10700 km/s] range
containing 232 galaxies, with a BWT mean velocity of 9769 km/s, and a
BWT velocity dispersion of 518 km/s; note that this velocity structure
is not quite centered on the velocity of the cD galaxy; a smaller peak
at 10940 km/s with 36 galaxies in the [10700, 11860 km/s] range.  If
we only keep the largest scales, we are left with a rather symmetric
velocity distribution showing an excess at high velocities.  This
excess corresponds to the peak at 10940 km/s, which contains a small
number of galaxies (see section 4.4).

\begin{figure*}[t]
\centerline{\psfig{figure=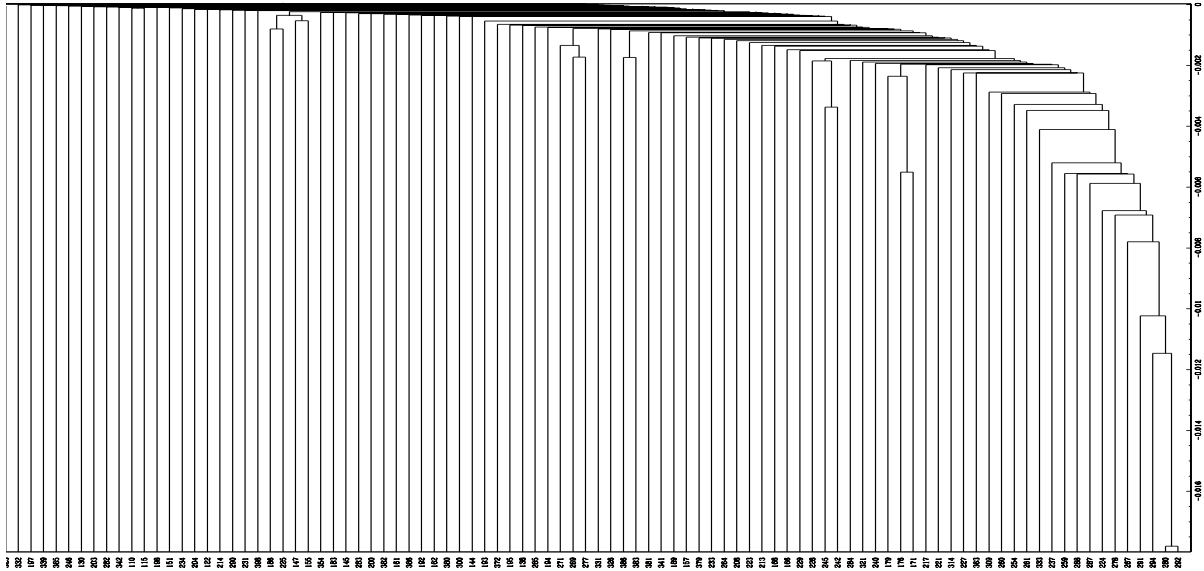,height=16cm,width=8cm,angle=90,clip=true}}
\caption[ ]{Dendogram obtained by applying a Serna \& Gerbal analysis to 
the subsample of 96 galaxies located within a radius of 1800 arcsec 
around the cD and with magnitudes R$\leq$17.0. The small numbers at the bottom
correspond to the galaxy numbers in our velocity catalogue. }
\protect\label{sernaamas}
\end{figure*}

These structures are also found by applying a rank-velocity
classification, which gives two breaks globally consistent with those
found by analyzing the cluster velocity distribution. Such breaks
probably indicate substructures with velocities coherent with the
finer analysis based on the wavelet technique. However, the number of
galaxies involved in these structures is small, and the velocity
distribution in the main cluster therefore appears to be quite smooth,
suggesting that Abell 496 is quite well relaxed.

In order to confirm the state of relaxation of Abell 496, we have
applied a Serna \& Gerbal (1996) analysis to the subsample of 96
galaxies located within a radius of 1800 arcsec around the cD and with
magnitudes R$\leq$17.0; within this limited sample, the completeness
of the redshift catalogue is 82\% and this type of analysis is
expected to give robust results. Note that galaxies in this region
with measured velocities but without published magnitudes were
discarded. Results are displayed in Fig. \ref{sernaamas}. At the
extreme lower right of the figure, we can see the very tight pair made
by the cD (\#280, R=12.2) and a satellite galaxy (\#292, R=15.6, in
the Durret et al. 1999b catalogue): this confirms that the cD is at
the bottom of the cluster potential well. We also observe a structure
of 11 bright galaxies (10 galaxies with R$\leq$15.8, plus one with
R=16.8) highly concentrated in space around the cD (mean distance to
the cD: 216 arcsec, with a dispersion of 48 arcsec) but not in
velocity (BWT mean velocity and velocity dispersion: 9745 and 375
km/s). This result is comparable to what is found in other clusters,
where the central core is more or less well discriminated.  The main
body of the cluster center appears quite relaxed, with no strong
subclustering within a radius of 1800 arcsec (1.65 Mpc). This picture
agrees with the general shape of \a4\ seen in X-rays (see
Fig. \ref{image}).

\subsection{Luminosity segregation in \a4}

After violent relaxation, two-body gravitational interactions lead to
a certain level of energy equipartition between galaxies of various
masses, and consequently to a certain segregation in velocity
dispersion with luminosity (mass). This process concerns essentially
massive galaxies, and added to dynamical friction, it creates
segregation with distance to the cluster center. The stage of
post-violent relaxation therefore leads to a segregation in the
[L,$\sigma _v$] space larger in the central regions than in the
overall cluster.

In order to search for such effects in \a4, we have derived the
velocity dispersion and the average distance of galaxies to the
cluster center (defined by the position of the cD) in several
magnitude bins.  We restrict our sample to galaxies belonging to the
cluster, i.e. in the velocity range [7813,11860 km/s], and within 1000
and 1800 arcsec from the cluster center, in order to have reasonably
complete samples:100\% and 79\% complete for R$\leq 18.5$
respectively. The completeness is estimated by comparing the number of
galaxies with measured redshifts to the number of galaxies in our
photographic plate catalogue, for the same R magnitude limit.  Since
there are 2 galaxies with R$< 14$ and 6 with 14$<$R$<$15, we chose to
fit the data with two different ``brightest'' bins: one including the
8 galaxies with 12$<$R$<$15, and the other including only the 6
galaxies with 14$<$R$<$15.

\begin{figure}[t!]
\centerline{\psfig{figure=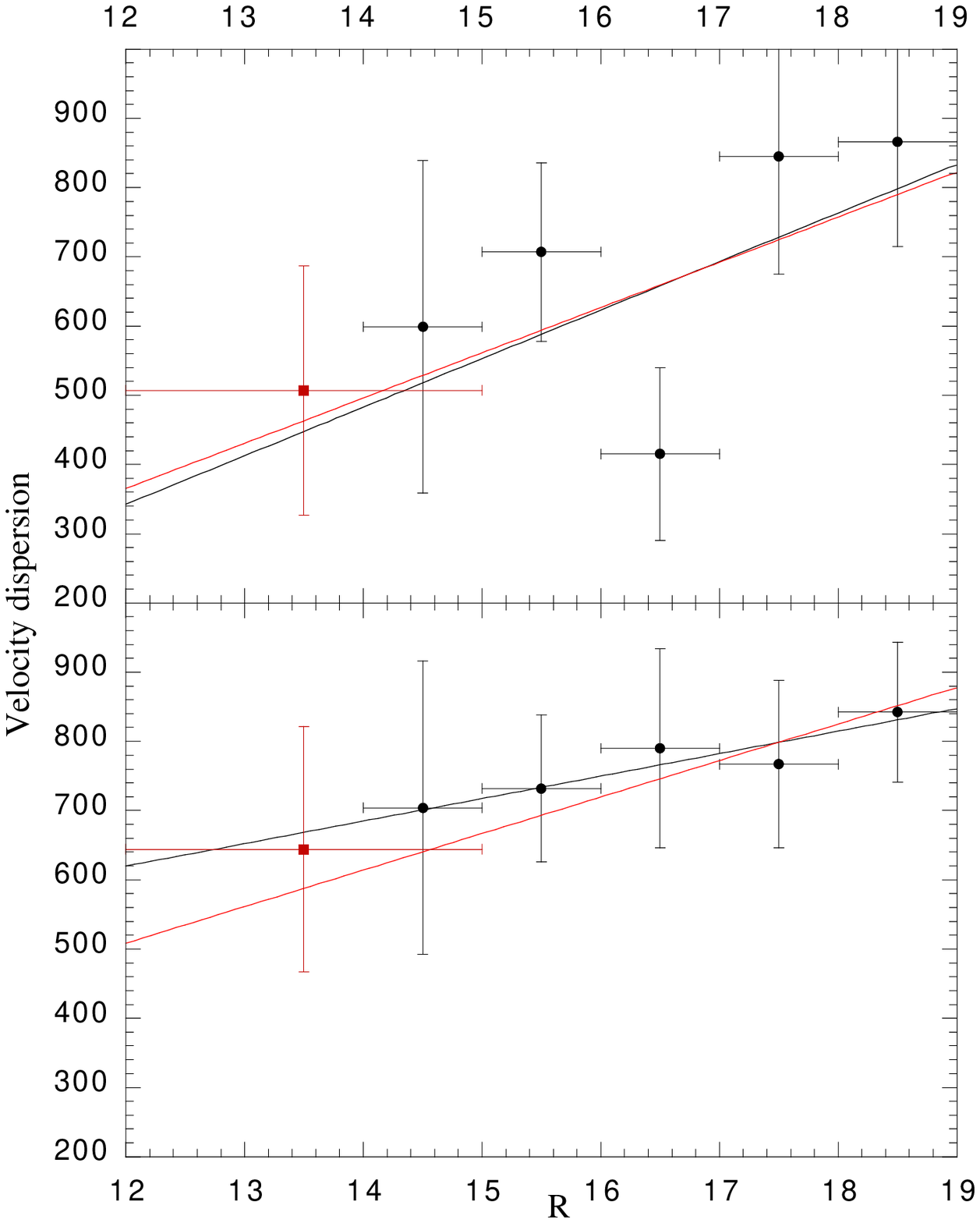,height=7cm}}
\caption[ ]{Velocity dispersion in several magnitude bins. The top
figure corresponds to a region of 1000 arcsec radius, the bottom figure
to a 1800 arcsec radius. The grey line is the fit including all the points,
while the dark line is the fit when the brightest bin is excluded.}
\protect\label{sigmavR}
\end{figure}

The velocity dispersions estimated in several magnitude bins are
different for the two samples, as shown in Fig. \ref{sigmavR}. In a
1000 arcsec radius, the velocity dispersion increases more steeply
with magnitude: the corresponding slopes are $65 \pm 43$ and $70 \pm
50$ km~s$^{-1}$~mag$^{-1}$ when the brightest bin is included or not
respectively.  In a 1800 arcsec radius, the velocity dispersion
increases less with magnitude, the corresponding slopes being $52 \pm
30$ and $32 \pm 41$ km~s$^{-1}$~mag$^{-1}$.

\begin{figure*}
\centerline{\psfig{figure=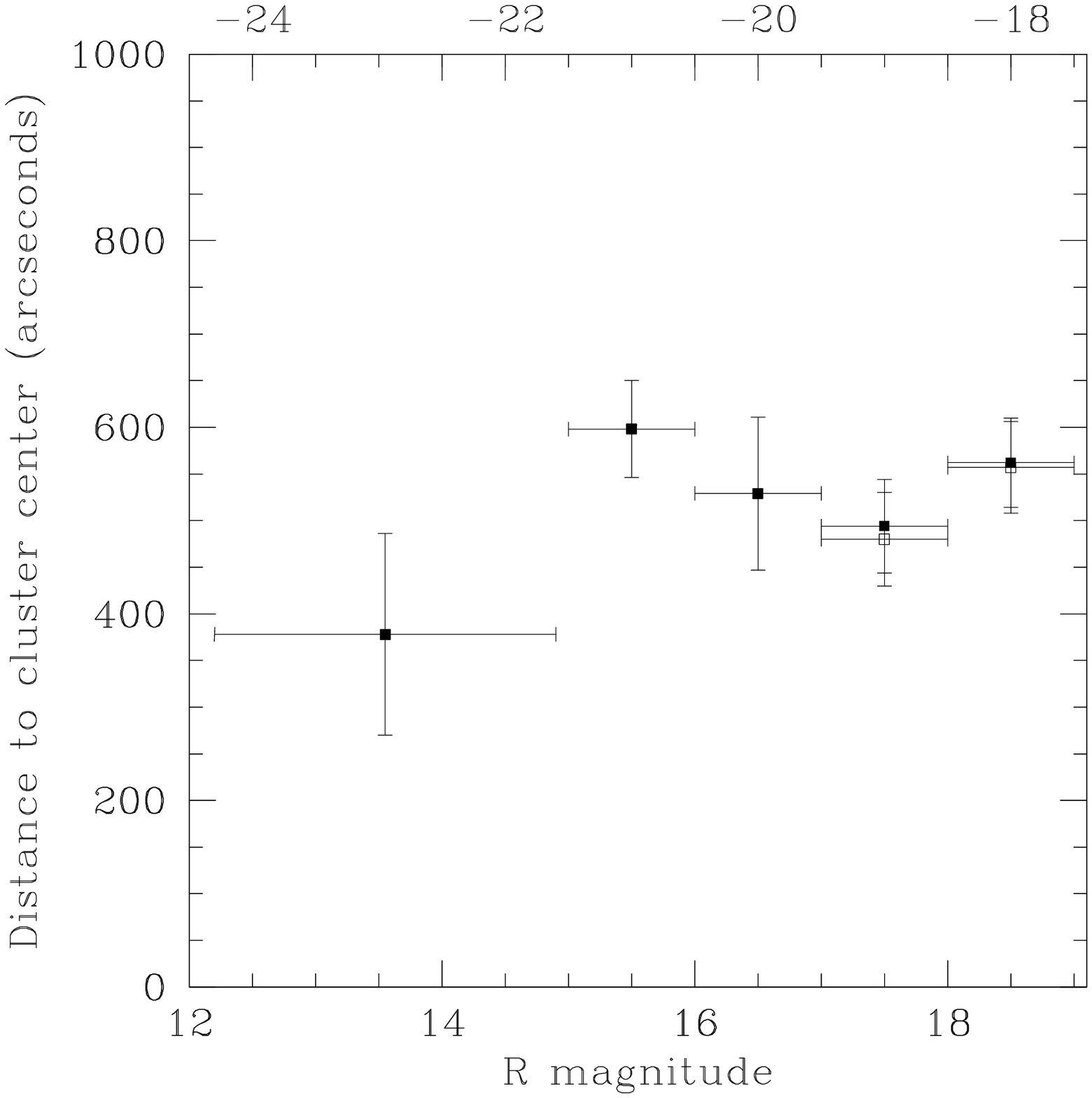,height=6cm}\qquad{\psfig{figure=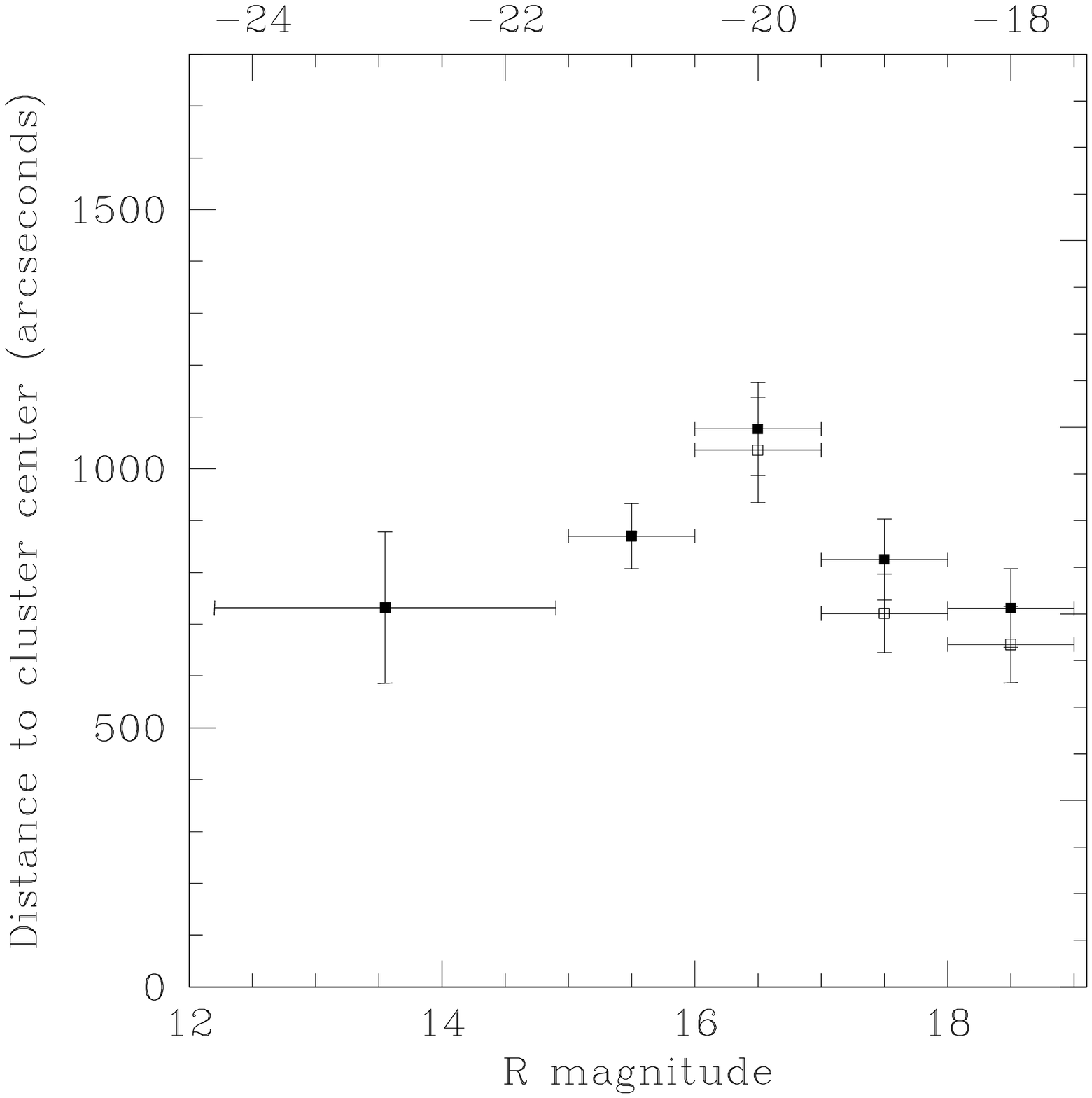,height=6cm}}}
\caption[ ]{Average distance to the cluster center for several
magnitude bins. The sample is restricted to galaxies in the velocity
range [7813,11860 km/s], within 1000 arcsec (left) and 1800 arcsec
(right) from the cluster center. Filled squares include all galaxies, 
empty squares indicate non-emission line galaxies.}
\protect\label{meandR}
\end{figure*}

As seen in Fig. \ref{meandR}, the average distance to the cluster
center is somewhat smaller for the galaxies located in the brightest
bin (R$\leq$15), then remains roughly constant with a possible
decrease with increasing magnitude, specially in the broadest sample
(1800 arcsec radius). 

The combination of Figs. \ref{sigmavR} and \ref{meandR} seems to
correspond well to a post-violent relaxation stage.

Interestingly, we can notice that both the distance to the cluster
center and the overall velocity dispersion range are reduced when
emission line galaxies (hereafter ELGs) are excluded.  This agrees
with the general scheme that ELGs are often found in the outskirts of
clusters of clusters and are not as strongly tied to the cluster
gravitationally (e.g. Biviano et al. 1997). We now discuss in more
detail the properties of ELGs in \a4.

\subsection{The emission line galaxy distribution}

\begin{figure*}
\centerline{\psfig{figure=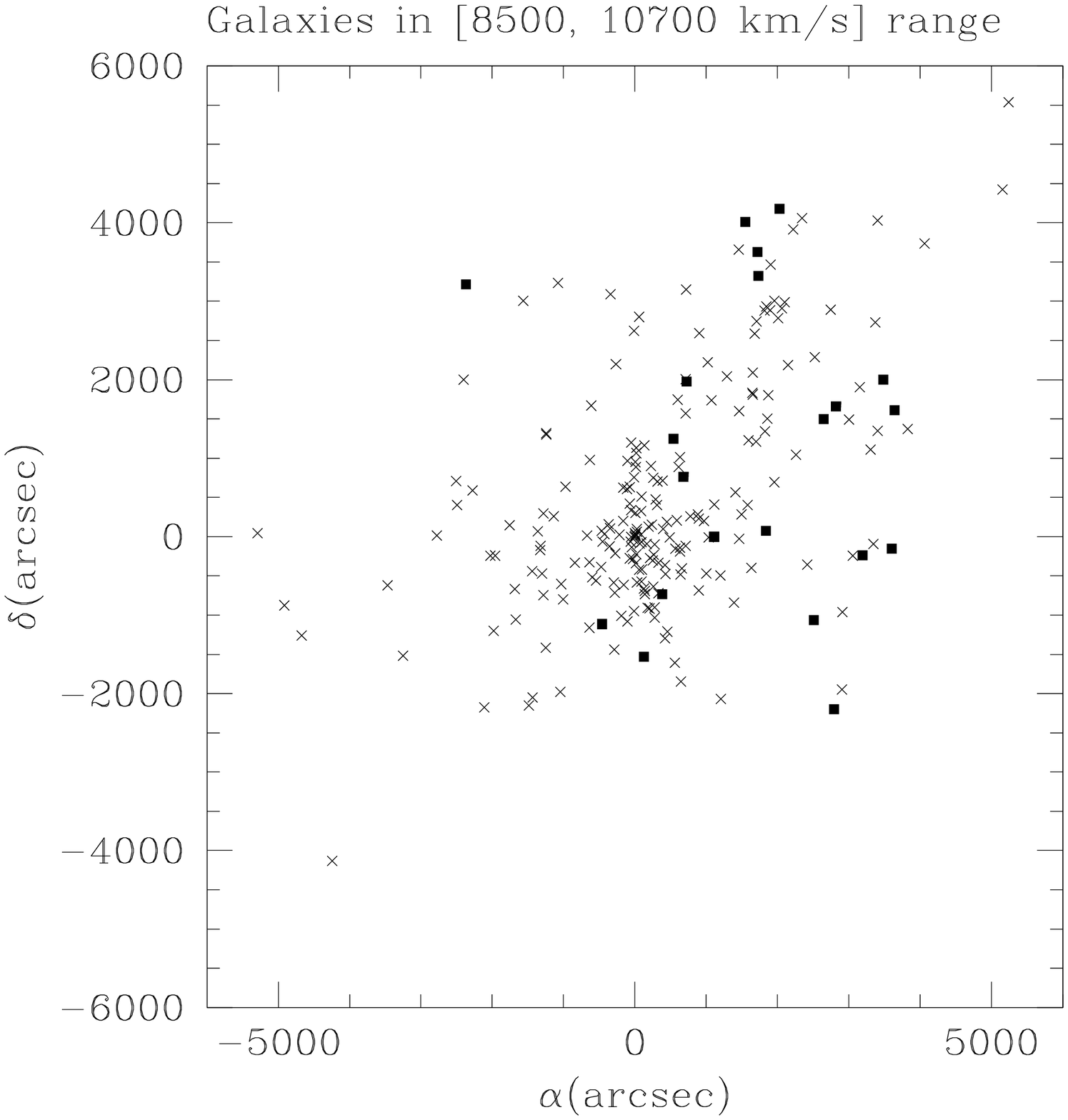,height=6cm}\qquad{\psfig{figure=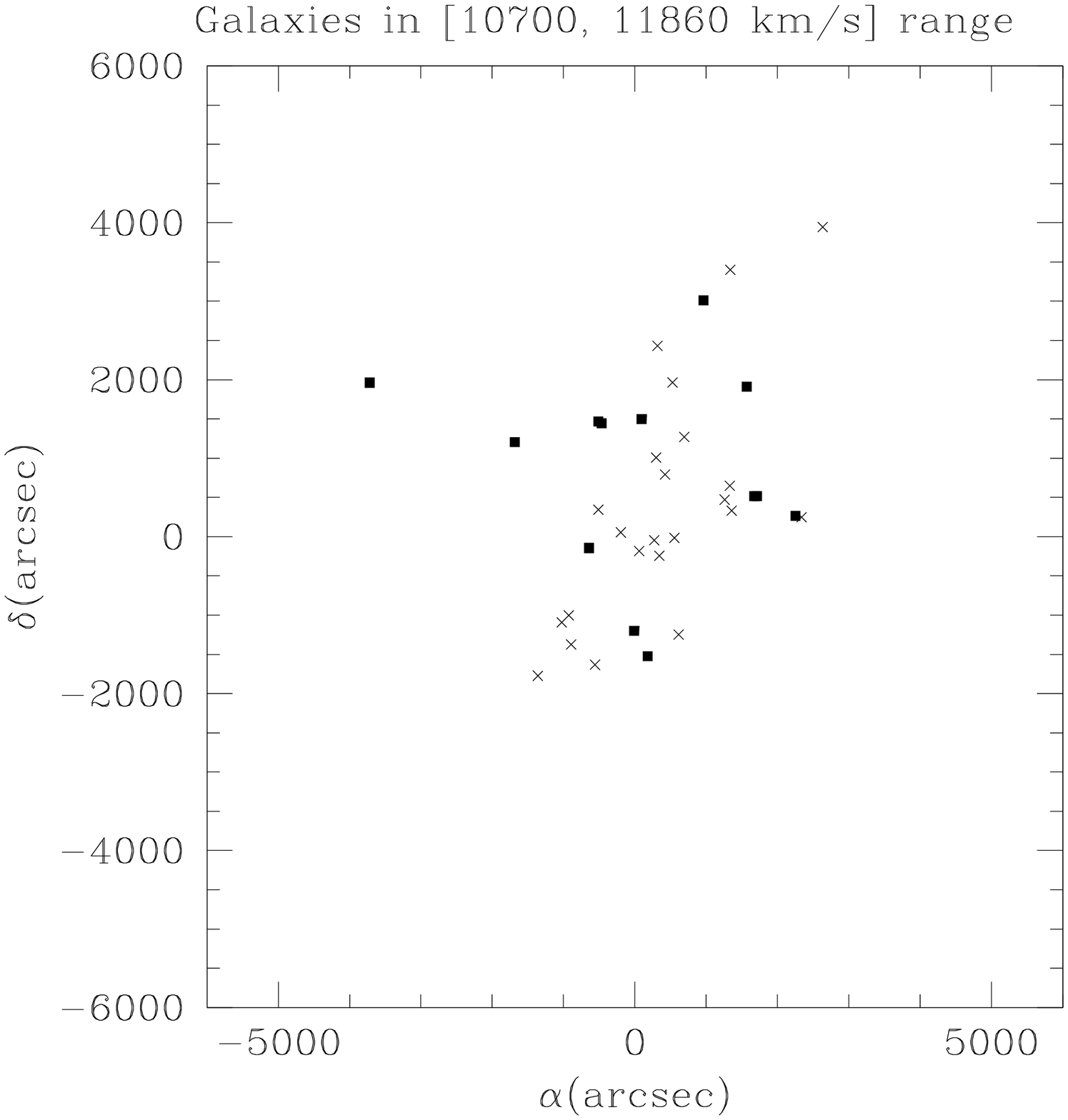,height=6cm}}}
\caption[ ]{Left: positions of the 211 non-emission line galaxies (crosses)
and of the 21 emission line galaxies (filled squares) in the 
[8500, 10700 km/s] velocity range. Right: positions of the 23 non-emission 
line galaxies (crosses) and of the 13 emission line galaxies (filled squares) 
in the [10700, 11860 km/s] velocity range.}
\protect\label{xymain}
\end{figure*}

We now compare the distribution of emission line (ELGs) versus
non-emission line (NoELGs) galaxies. There are 85 ELGs and 381 NoELGs
in our velocity catalogue, among which 34 ELGs and 241 NoELGs in the
velocity range of \a4. The global percentage of ELGs in the cluster is
therefore $\sim 12 \pm 3$\%. Note that this percentage is perfectly
coherent with the proportions observed by Biviano et al. (1997) in the
ENACS survey.

The spatial distribution of the 211 NoELGs and 21 ELGs belonging to
the main [8500, 10700 km/s] velocity peak is displayed in
Fig. \ref{xymain}. The fraction of ELGs in this velocity range is
9$\pm$3\%, consistent with the global cluster value within the error
bars. NoELGs appear rather homogeneously distributed, except for a
sort of linear north-south concentration towards the center. On the
other hand, a large majority of the ELGs in this velocity range is
located west of a north to south line crossing the center, and at
least half of these ELGs even seem to be close to the west cluster
edge. This agrees with the fact that ELGs tend to concentrate in the
outer regions of clusters (see e.g. results based on ENACS data by
Biviano et al. 1997). The presence of an excess of ELGs can at least
in some cases be interpreted as due to merging events producing shocks
which trigger star formation. This was shown to be the case in the
zone of Abell 85 where the X-ray filament merges into the main body of
the cluster (Durret et al. 1998): an excess of ELGs was observed in
that region, together with a temperature increase of the X-ray gas.
The ASCA X-ray gas temperature map presently available for Abell 496
(Markevitch et al. 1999, Donnelly, private communication), does not
show any temperature increase for the X-ray gas in that area, so we
cannot correlate the excess of ELGs towards the west cluster edge with
a higher gas temperature zone. However, we can note that this excess
is located roughly in the same region as the X-ray excess emission in
the north west region of the cluster (Fig. \ref{image}). Such an X-ray
enhancement could be due to a merging event originating from the north
west, but our data cannot show this with certainty.

On the other hand, the spatial distributions of the 23 NoELGs and 13
ELGs in the [10700, 11860 km/s] velocity interval are comparable
(Fig. \ref{xymain}), while the ELG fraction seems much higher:
36$\pm$18\%. Though the small number of objects may introduce errors,
there definitely seems to be an excess of ELGs with somewhat higher
velocities than the bulk of the cluster; these ELGs account at least
partly for the peak at 10940 km s$^{-1}$ in the wavelet reconstruction
of the velocity distribution.

A general picture for the ELG distribution in \a4\ is that of two
samples of galaxies falling on to the main cluster, one from the back
(the ELGs concentrated towards the west) and one from the front (the
high velocity ELGs). 

\subsection{The galaxy luminosity function}

We have seen in the previous section that \a4\ appears to have
properties common to many clusters, with a relaxed main body and ELGs
probably falling on to the cluster.  We therefore expect its galaxy
luminosity function not to be strongly modified by environmental
effects, as observed in some clusters showing more prominent
substructures. We discuss below its main features.

\subsubsection{The bright end of the galaxy luminosity function}

\begin{figure*}[t!]
\centerline{\psfig{figure=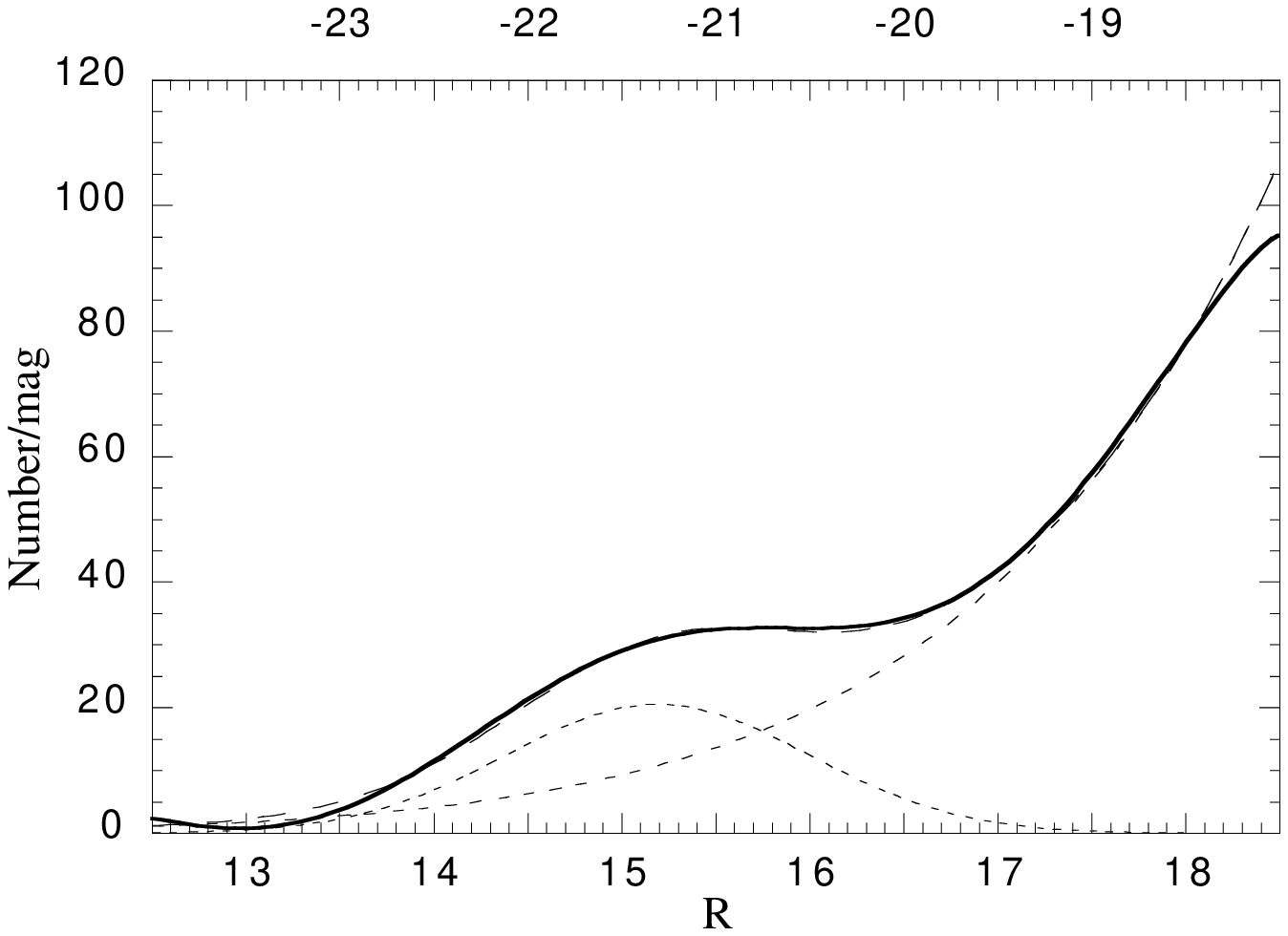,height=6cm}\qquad{\psfig{figure=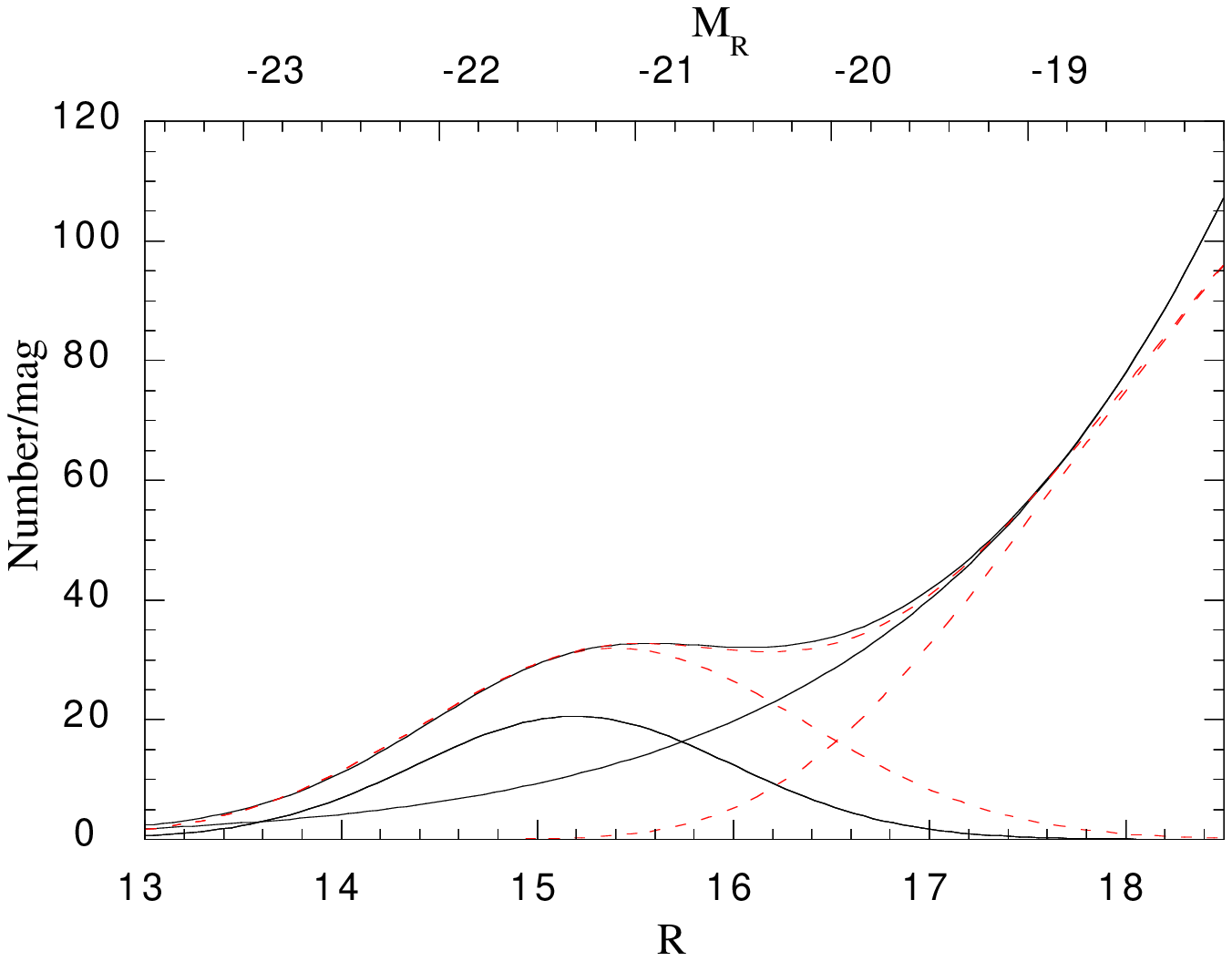,height=8cm,angle=90,clip=true}}}
\caption[ ]{Left: wavelet reconstruction of the luminosity function in
the R band for the 196 galaxies in the redshift catalogue with
velocities in the cluster range, within a radius of 1800 arcsec and a
limiting magnitude R=18.5 (full line).  The contribution of the bright
and faint populations accounted for by a gaussian and a power law
respectively are also shown (dots and short dashes), as well as their
sum (long dashes). The top scale indicates R absolute
magnitudes. Right: contributions of the bright and faint populations
accounted for by a gaussian and a power law (full line), and by a
gaussian and a Schechter function (dashes). The sums of both functions
are also displayed. }
\protect\label{Rredshifts}
\end{figure*}

We have first derived the galaxy luminosity function (GLF) of \a4\ in
the R band from the redshift catalogue, within a radius of 1800 arcsec
around the center, and for a limiting magnitude R=18.5. There are 196
galaxies in this sample.  The completeness of the redshift catalogue
within these limits is 79\%, and it is 100\% in that region for
R$\leq$16.0. The obvious interest of such a GLF is that no background
contribution needs to be subtracted, therefore making the results very
robust. We have limited the magnitude interval to the [13,18.5] range,
because for R$\leq$13 there is only one galaxy (the cD), which
introduces edge effects in the wavelet reconstruction of the GLF, and
for R$\geq$18.5 the completeness sharply decreases. This corresponds
to the [$-23.5,-18.0$] interval in R absolute magnitude.

The GLF obtained after a wavelet reconstruction is shown in Fig.
\ref{Rredshifts}.  The sample was analyzed with 128 points, excluding
the two smallest scales. The significance level of the detected
features is at least 3$\sigma$. A flattening is observed at R$\sim$16,
corresponding to an absolute magnitude M$_{\rm R}\sim -20.5$. This
shape is comparable to that found in Virgo (Binggeli et al. 1988) and
in Abell 963 (Driver et al. 1994), where a comparable flattening was
observed at a common absolute magnitude of $-19.8$. The GLFs of Coma
and Abell 85 are more complex, with a ``bump'' corresponding to the
brightest galaxies, followed by a ``dip'' at M$_{\rm R}\sim -20.5$
(see Fig. 9 in Durret et al. 1999a).  Note that the flattening of the
GLF in \a4\ occurs exactly at the same absolute magnitude as the dip
in Abell 85 and Coma.

The GLF in \a4\ suggests at least a bimodal galaxy distribution, with
bright (mostly elliptical) galaxies in the bright part and dwarf
galaxies in the fainter part. We therefore performed a fit of the
wavelet reconstructed GLF of \a4\ by summing two functions: a gaussian, to
account for bright galaxies, and a power law (case 1) or a Schechter
function (case 2) to represent faint and/or dwarf galaxies. In case 1,
we fit the data as a function of apparent R magnitude; the gaussian is
then found to be centered on R=15.19$\pm$0.01, with $\sigma = 0.8 \pm
0.1$, and the power law varies as R$^{11.67 \pm 0.16}$. In case 2, we
fit the data as a function of absolute R magnitude, to allow a direct
comparison with other authors; the gaussian is then found to be a
little broader, centered on R=15.48$\pm$0.04, with $\sigma = 1.0 \pm
0.1$; the Schechter function, defined as in Rauzy et al. (1988,
section 3.2.3), has $\alpha = -1.19\pm 0.04$ and M$_* = -19.43 \pm 0.13$
(in the [$-23.5,-18.0$] absolute magnitude range).

The GLF resulting from these various fits is shown in
Fig. \ref{Rredshifts}; it obviously reproduces the data very
well. Except at the faint end where the sample incompleteness most
probably modifies the GLF shape, both fits 1 and 2 are good but we
cannot distinguish between them. In view of the obvious quality of the
fit, we did not attempt to estimate error bars with Monte-Carlo
simulations, as done previously e.g. for Abell 85 (see Durret et al.
1999a, Fig. 12).

The various values obtained from these fits of the GLF can be compared
to those found in other clusters. The gaussian used to fit the bright
part of the GLF in Abell 85 has $\sigma =1.1$, comparable to the value
we find in case 1. The Schechter function for \a4\ has a slope
comparable to that found by Lumsden et al. (1997), but notably flatter
than the values found in other surveys (e.g. Valotto et al. 1997,
Rauzy et al. 1998 and references therein). Rauzy et al. (1998) argued
that the flatter slope found by Lumsden was due to incompleteness at
faint magnitudes; this may also be true in our case, since our sample
is 100\% complete only to R=16.0 (M$_{\rm R}=-20.5$), and we also find
a brighter value of M$_*$ than the above surveys, suggesting that we
are missing faint galaxies.

We can note that the GLFs of Coma and Abell 85 were interpreted in a
similar way, with the bright part mainly due to ellipticals (with a
small contribution of spirals) and the faint part due to dwarfs
(Durret et al. 1999a). Comparable shapes were found in several other
clusters.  The fact that the GLF of \a4\ shows a flattening at the
same value M$_{\rm R}=-20.5$ indicates that the galaxy population in
\a4\ is comparable to those of the above mentioned clusters.

Note that Molinari et al. (1998) have analyzed the GLF of \a4\ from a
photometric catalogue in three colors, reaching magnitudes much
fainter than those of our spectroscopic catalogue. We will therefore
compare our results to theirs in the next section.

\subsubsection{The faint end of the galaxy luminosity function}

\begin{figure}
\centerline{\psfig{figure=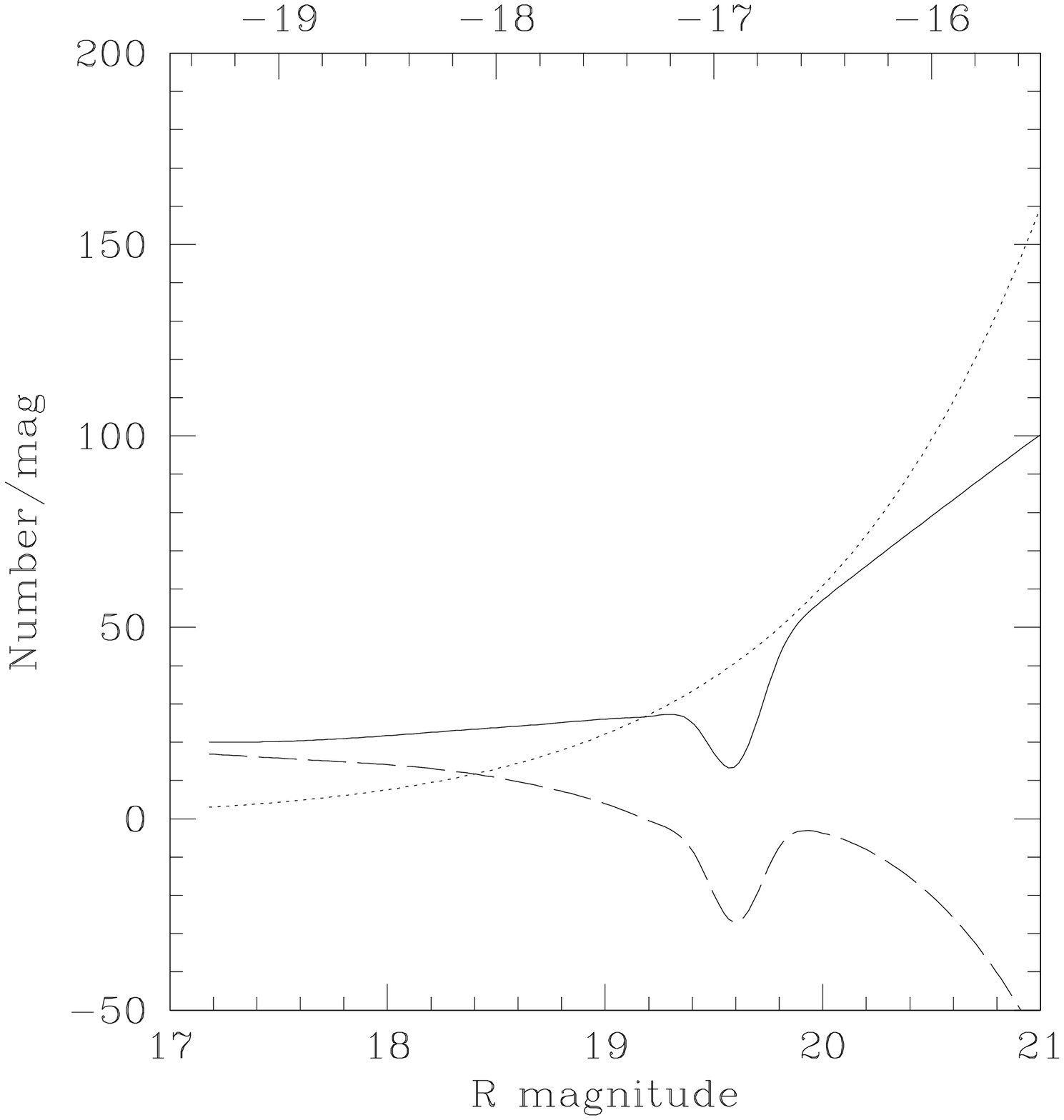,height=7cm}}
\caption[ ]{Wavelet reconstruction of the number of galaxies in the R band 
for the 411 galaxies in the CCD catalogue within the R magnitude range
[17,21] (full line). The background contribution obtained from the LCRS and 
ESS is shown as a dotted line (see text). The difference between the 
observed number of galaxies and the background is shown as a long dashed 
line. }
\protect\label{Rccd}
\end{figure}

Our intent was also to derive the luminosity function from the CCD
catalogue, which corresponds to a small region of $\sim$246 arcmin$^2$
in the cluster center. For this we first performed a wavelet
reconstruction of the R magnitude distribution in the R magnitude
range [17,23]. Since we have no background exposure, we estimated the
background contribution by connecting the counts from the Las Campanas
Redshift Survey (LCRS, Lin et al. 1996) and from the ESO-Sculptor
Survey (ESS, Arnouts et al. 1997), as described in our study of Abell
85 (Durret et al. 1999a, Fig. 10 and text), and we subtracted this
background to the observed number of galaxies. The result is shown in
Fig. \ref{Rccd}.  We have checked that the consistency of the
background number counts estimated by Tyson (1988) with those of the
LCRS and ESS combined as described above is good.

The difference between the observed number of galaxies and the
background (Fig. \ref{Rccd}) becomes negative for magnitudes R$\geq
18.4$, while the CCD catalogue is complete at least up to
R=21. Therefore, this background cannot be considered as
representative of the local background in our CCD field of view.  Note
that this was already the case for the CCD photometric data of Abell
85.

One notable feature is the dip in the galaxy magnitude distribution at
R$\sim 19.5$ (M$_{\rm R} \sim -17$), which is detected at a high
confidence level. This dip corresponds to that observed by Molinari et
al. (1998), who found a dip at R$\sim$19 (M$_{\rm R} \sim
-17.5$). Note that they also find a similar dip in the g band, and
possibly in the i band.  Molinari et al. (1998) made a second
determination of the GLF by selecting cluster members in a
colour-magnitude diagram. In this case, they find a small dip, or at
least a flattening, for R$\sim 18$ (M$_{\rm R} \sim -18.5$). This
value does not agree either with the bright nor with the faint GLF
that we derived. It is difficult to understand why, since their
colour-magnitude relation appears quite well defined.

\begin{figure}
\centerline{\psfig{figure=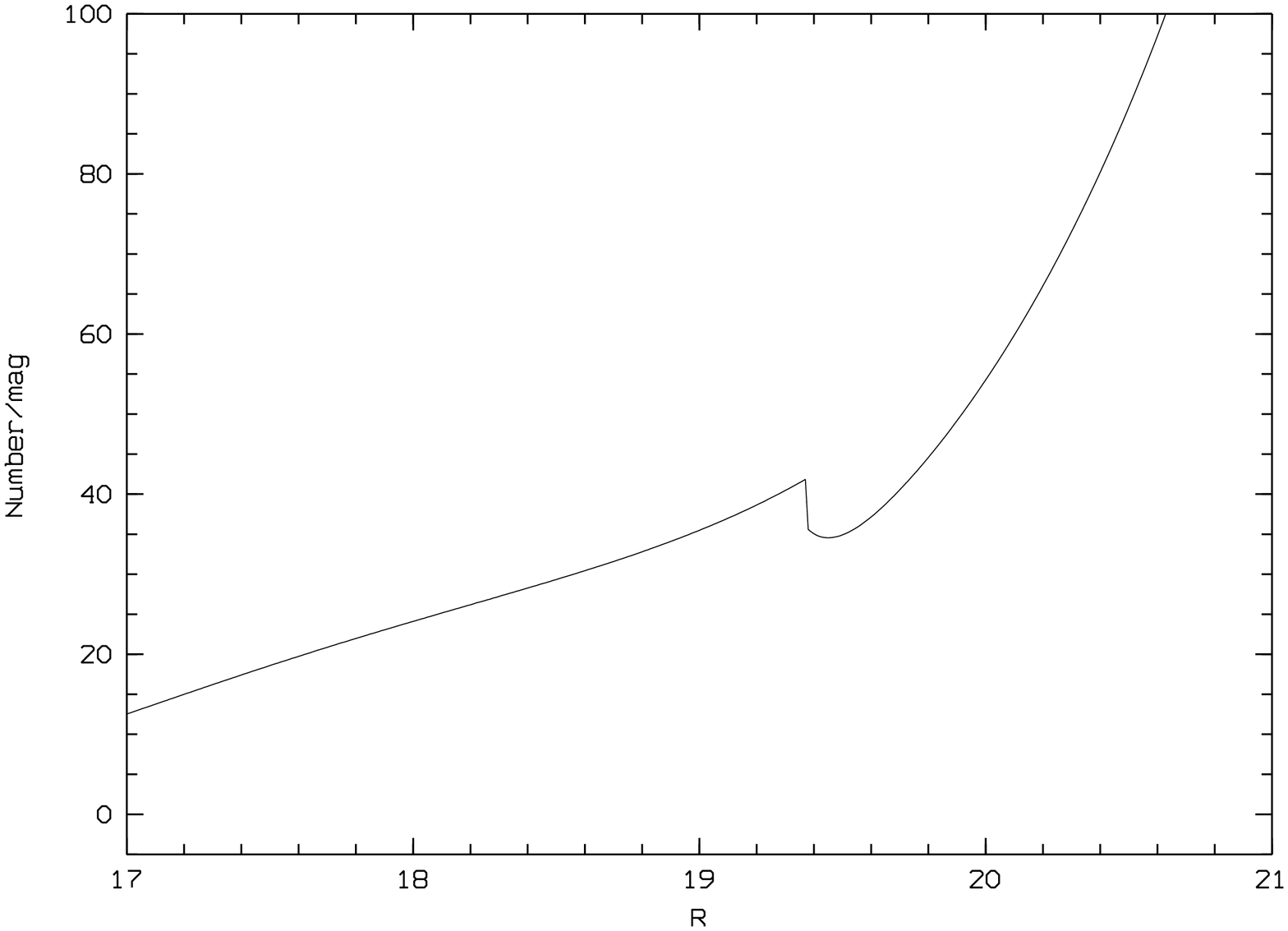,height=7cm,angle=-90}}
\caption[ ]{Model of the total counts reproducing the dip observed in Fig.
\ref{Rccd} (see description in text).}
\protect\label{fdlada}
\end{figure}

In order to investigate the origin of the dip seen in our data, we
propose a toy model, which is not a fit but only illustrates how the
dip could be accounted for. Let us first note that the contribution of
the other structures detected along the line of sight is negligible.
Assuming a Gaussian + a Schechter function to model the GLF (see
section 4.5.1), we rescaled the number of galaxies produced by this
composite function to fit the dimension of the CCD field.  We then
applied a magnitude cut-off to this GLF, as suggested by Adami et
al. (2000), for galaxies fainter than M$_{\rm R}=-19.75$ in the inner
core of the Coma cluster. This effect becomes very strong for galaxies
fainter than M$_{\rm R} =-17$. The exact shape of such a cut-off is
unknown, so we applied a convenient apodization function (the choice
of this function influences the shape and smoothness of the dip). The
background counts were modeled as the background contribution from the
LCRS and ESS described above.  We then summed the cluster and
background contributions, and the result is shown in
Fig. \ref{fdlada}.

Such a toy model can reproduce the global GLF shape, with counts
similar to the observed data and a dip comparable to the observed one.
A fine-tuning of the various parameters involved could make
Figs. \ref{Rccd} and \ref{fdlada} more similar, but this would push
the model too far. However, we can state that a cut-off in the GLF of
\a4\ similar to that observed in Coma is a solution to account for the
observed dip.

\section{The X-ray gas}

\begin{figure}[h]
\centerline{\psfig{figure=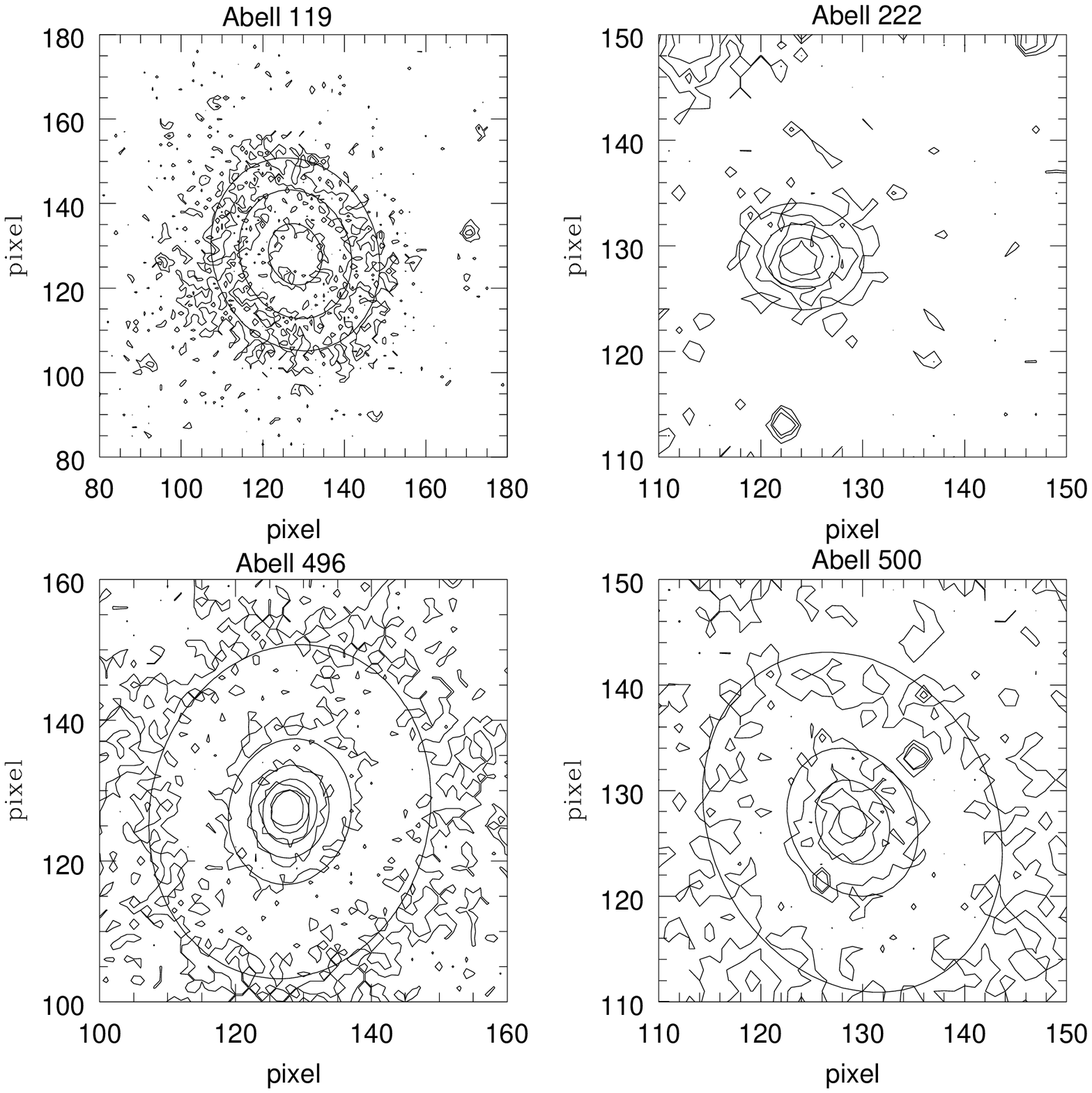,height=7cm,width=7.5cm,clip=true}}
\caption[ ]{Observed ROSAT PSPC image with the isocontours of the best 
$\beta$-model fit superimposed (see text). The pixel size is 30 arcsec.}
\protect\label{fitx}
\end{figure}

A pixel by pixel fit was performed on the X-ray image, as described by
Pislar et al. (1997). The pixel size is 30 arcsec. A $\beta$-model and
a 3D S\'ersic model (Lima Neto et al. 1999) were considered for the
variations of the density with radius.  The global temperature
estimated from these ROSAT data, using a Raymond-Smith spectrum and a
Galactic absorption column density was found to be 4$\pm$1 keV and
assumed to be constant (Pislar 1998). This is consistent with the
temperatures of 3.9 and 4.7 keV previously measured with the Einstein
and EXOSAT satellites respectively (David et al. 1993; Edge \& Stewart
1991). The parameters corresponding to the best fits for both models
are given in Table \ref{tabfitx}, and the result of the $\beta$-model
1 fit superimposed on the observed image is displayed in
Fig. \ref{fitx}.  We observe that in model~2 the central density is
lower than in model~1 and that the $\beta$ and r$_c$ parameters are
higher. This is because in model 2 we do not include the central
region, where the cooling flow lies.  The effect is the same for
models 3 and 4.

\begin{table}
\caption{Fits of the X-ray gas with a $\beta$-model and a S\'ersic model}
\begin{tabular}{lrrrrrr}
\hline
Model & n$_0$ & $\beta$ & r$_c$ & $\epsilon$ & R$_{cool}$ & $\dot{M}$\\
      &(10$^{-3}$ cm$^{-3}$) & & (kpc) & & (kpc) & (M$_\odot$/yr)\\
\hline
1     & 21.7 & 0.53 & 50 & 0.87 & 192 & 235 \\
      &  1.9 & 0.01 &  4 & 0.03 &  25 &  81 \\
2     &  2.55 & 0.82 & 356 & 0.86 & & \\
      &  0.35 & 0.07 &  58 & 0.03 & & \\
\hline
Model & I$_0$ & $\nu$ & a & $\epsilon$ & R$_{cool}$ & $\dot{M}$\\
      &(10$^{-3}$ cm$^{-3}$) & & (kpc) & & (kpc) & (M$_\odot$/yr) \\
\hline
3     & 5.5 & 0.61 & 230 &0.86 &198 & 314  \\
      & 0.3 & 0.02 &  12 & 0.03 &   &  \\
4     & 2.7 & 0.75 & 368 & 0.85 &   & \\
      & 1.3 & 0.17 & 129 &  0.03 &  & \\
\hline
\end{tabular}

Notes to Table \ref{tabfitx}: n$_0$, $\beta$ and r$_c$ are the parameters of
the best $\beta$-model fit, and I$_0$, $\nu$ and a those of the best
S\'ersic fit; $\epsilon$ is the cluster ellipticity, R$_{cool}$ the
cooling radius and  $\dot{M}$ the cooling flow mass deposit rate.\\
Models 1 and 2 correspond to a $\beta$-model, 
with the central region respectively included or not; models 3 and 4 
correspond to a S\'ersic model with the central region respectively 
included or not. For each model, the first line gives the parameters and the
second line the corresponding 3$\sigma$ errors.
\protect\label{tabfitx}
\end{table}

Our values of $\beta$ and r$_c$ (in model~2) are higher than those of
Markevitch et al. (1999), who found $\beta=0.7$ and
r$_c=249$~kpc. This is due to the fact that they exclude a central
region smaller than ours (3 arcmin instead of 3.3 arcmin).  Pislar
(1998) has shown that in a cooling flow cluster the bigger the
excluded central region, the higher $\beta$ and r$_c$, and the lower
the central density. Moreover, at the image limiting radius, the
values of the dynamical and gas masses do not depend on the size of
the central region excluded.

The cooling radius R$_{cool}$ and the mass $\dot{M}$ deposited in the
centre were estimated as in Pislar et al. (1997) and are given in
Table \ref{tabfitx} for a temperature of 4 keV. We take $2\,10^{10}$
yr for the cooling time.

\begin{figure}
\centerline{\psfig{figure=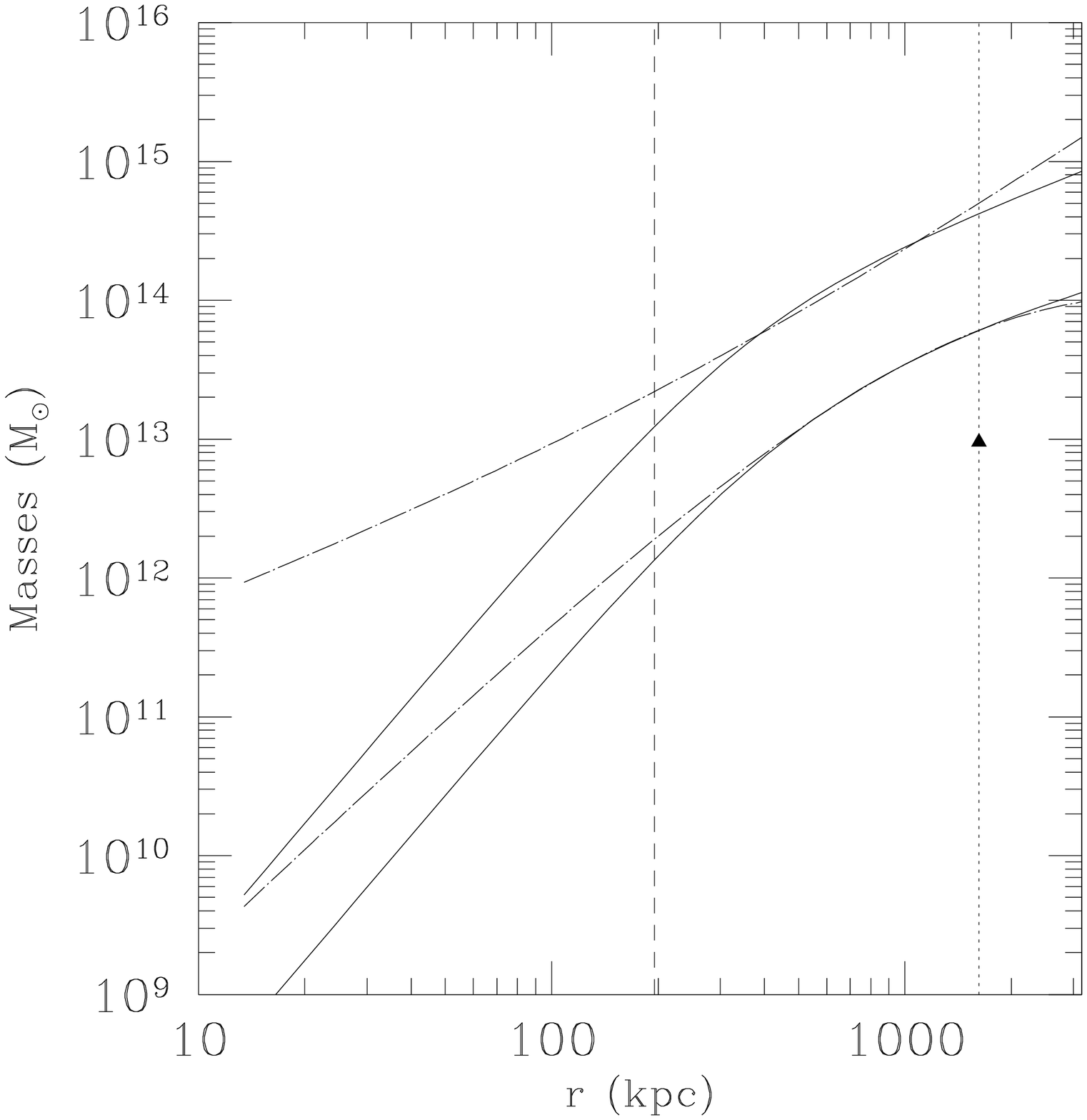,height=7cm}}
\caption[ ]{Masses of the X-ray gas (lower curves) and total dynamical 
masses (upper curves) derived from the X-ray data. Full lines correspond to
$\beta$-model fits and dot-dashed lines to S\'ersic. The vertical
lines indicate the cooling radius (dashes) and the limiting radius of the 
image (dots). The filled triangle indicates the integrated galaxy mass at
the X-ray limiting radius.}
\protect\label{masses}
\end{figure}

\begin{figure}
\centerline{\psfig{figure=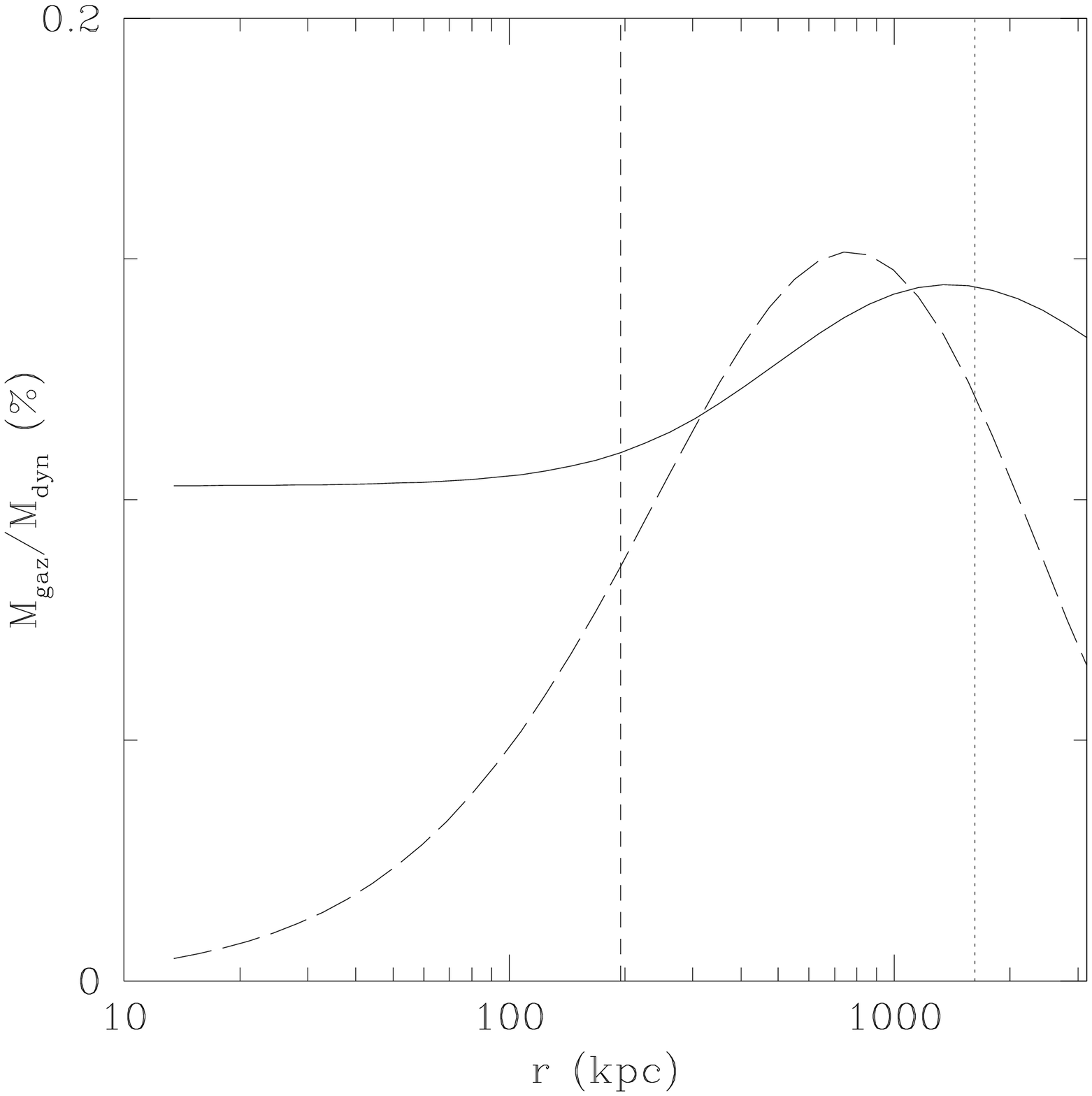,height=7cm,clip=true}}
\caption[ ]{X-ray gas to dynamical mass ratio. The full line corresponds 
to the $\beta$-model fit and the dot-dashed line to the S\'ersic fit.
The vertical lines indicate the cooling radius (dashes) and the limiting 
radius of the image (dots).}
\protect\label{baryons}
\end{figure}

The X-ray gas and total dynamical masses derived from the X-ray data
assuming equilibrium are shown in Fig. \ref{masses} as a function of
radius. Note that these curves are only valid between the cooling
radius and the limiting radius of the X-ray image; within this
validity range, both models are in good agreement with each other.

We find, with models 2 and 4 for a prolate geometry, at the limiting
radius of the image, a gas mass of $(6.1\pm2.2)\ 10^{13} M_{\odot}$
and a dynamical mass of $(4.2\pm1.1)\,10^{14} M_{\odot}$. We
overestimate the errors because we have supposed that the parameters
are not correlated. The masses at $1\,h_{50}^{-1}$ Mpc are
respectively $(3.45\pm1.1)\,10^{13} M_{\odot}$ and
$(2.4\pm0.6)\,10^{14} M_{\odot}$. The gas mass found by Mohr et
al. (1999) and the dynamical mass found by Markevitch et al. (1999)
are very similar if we remember that their geometry is spherical.  The
mass calculated by the virial theorem applied to all the galaxies in
the redshift catalogue with velocities in the cluster is M$_{vir}=(7.2
\pm 0.8) 10^{14}$ M$_\odot$. This value agrees with the dynamical mass
derived above, providing the mass profiles can be extrapolated to
radii larger than the X-ray image. The integrated mass of galaxies
within the X-ray image limiting radius, assuming a mass to luminosity
ratio of 8 M$_\odot$/L$_\odot$, is $1.0\ 10^{13}$ M$_\odot$. Note
that, due to incompleteness of our redshift catalogue, in particular
at faint magnitudes, this is only a lower limit.

The ratio of the X-ray gas mass to the dynamical mass is shown as a
function of radius in Fig. \ref{baryons}. The fraction of X-ray gas is
about 0.15 at the limiting radius of the image and at $1\,h_{50}^{-1}$
Mpc with a $\beta$-model. With a S\'ersic model, the ratio is 0.12 at
the limiting radius of the image. The baryon fraction that we find at
$1\,h_{50}^{-1}$ Mpc with the $\beta$-model is very similar to that
obtained by Markevitch et al. (1999) ($0.158\pm0.017$), and comparable
to that found in other clusters. At the X-ray limiting radius, the
stellar to X-ray gas mass ratio is 16\% and the stellar to total
dynamical mass ratio is 2.4\%.

\section{Discussion and conclusions}

The optical analysis of the \a4\ field has shown the existence of
several structures along the line of sight. Among these, one
(structure 6) is likely to be a poor, diffuse and low mass cluster,
while two others (structures 4 and 9) are probably filaments more or
less aligned along the line of sight, the latter presenting a smooth
velocity gradient. Notice that the distances between these various
structures are very large.

The cluster \a4\ itself has quite a regular morphology. It includes
274 galaxies in the [7813,11860 km/s] velocity range and has a
velocity dispersion of 715 km/s. Its velocity distribution implies a
small amount of substructure. The analysis of the correlations between
position, luminosity and velocity dispersion indicates a post-violent
relaxation state. We can notice that both the distance to the cluster
center and the overall velocity dispersion ranges are reduced when
emission line galaxies (hereafter ELGs) are excluded.  This agrees
with the general scheme that ELGs are often found in the outskirts of
clusters and are not as strongly tied to the cluster
gravitationally (e.g. Biviano et al. 1997).  There may be two samples
of ELGs falling on to the main cluster, one from the back (the ELGs
concentrated towards the west) and one from the front (the high
velocity ELGs).

The bright luminosity function derived from our redshift catalogue
shows a flattening at R$\sim 16$ (M$_{\rm R}\sim -20.5$), comparable
to similar shapes found in other clusters. This suggests at least a
bimodal distribution, one for ellipticals and one for fainter
galaxies. The fact that the flattening occurs at the same absolute
magnitude as for other clusters suggests that the galaxy populations
in all these clusters are comparable.  

At fainter magnitudes, galaxy counts derived from CCD imaging show a
dip at R$\sim 19.5$ (M$_{\rm R} \sim -17$) which can be reproduced if
we assume a magnitude cut-off similar to that observed in Coma (Adami
et al. 2000). Notice that such a cut-off is observed in the very
central regions of both clusters. Although this result is only based
on imaging and remains to be confirmed spectroscopically, we may be
evidencing a second example of a cut-off of the faint end of the
luminosity function in a cluster.

We have modelled the X-ray gas and derived the X-ray gas mass and the
dynamical mass, which we compare to the stellar mass. At the limiting
radius (1.62 h$_{50}^{-1}$ Mpc) of the image, we find a fraction of
X-ray gas to total mass of 0.12$-$0.15 and a stellar to X-ray gas mass
ratio of 0.16.  We can note that \a4\ follows exactly the two by two
correlations between the X-ray luminosity (L$_{\rm X}=6.8\ 10^{44}$
erg/s, Wu et al. 1999), the X-ray temperature (T$_{\rm X}=4$ keV) and
the galaxy velocity dispersion ($\sigma _v = 715$ km/s) described in
the literature (see e.g. Wu et al. 1999 and references therein). These
values are typical of a richness class 1 cluster.

\a4\ therefore appears to be a relatively quiet and simple cluster,
with no strong environmental effects, although we may see an
enhancement of the X-ray emission and of the number of emission line
galaxies towards the north west.

While Coma has long been the archetype of a relaxed cluster and is
not believed to be relaxed any more (see Biviano 1998 and the
proceedings of the Coma meeting), the results presented above suggest
that \a4\ may be such a prototype, and can be used as a ``template''
in the future study of more complex (i.e. substructured) clusters.

\acknowledgements{The authors thank Andrea Biviano for help. C.~Adami
is grateful to the staff of the Dearborn Observatory for their
hospitality during his postdoctoral fellowship. We acknowledge
financial support from the French Programme National de Cosmologie,
CNRS.}


\begin{thebibliography}{}

\bibitem{} Abell G.O. 1958, ApJS 3, 211
\bibitem{} Adami C., Ulmer M., Durret F. et al. 2000, A\&A in press,
astro-ph/9910217
\bibitem {} Arnouts S., de Lapparent V., Mathez G. et al. 1997, A\&AS 124, 
163 
\bibitem{} Beers T.C., Flynn K., Gebhardt K. 1990, AJ 100, 32
\bibitem{} Binggeli B., Sandage A., Tammann G.A. 1988, ARA\&A 26, 509
\bibitem {} Biviano A., Katgert P., Mazure A. et al. 1997, A\&A 321, 84
\bibitem {} Biviano A. 1998, Proc. ``A new vision of an old cluster: 
untangling Coma Berenices'', Marseille June 17-20 1997, Eds. Mazure et al.,
World Scientific, page 1
\bibitem{} Condon J.J., Cotton W.D., Greisen E.W. et al. 1998, AJ 115, 1693
\bibitem{} Dantas C.C., de Carvalho R.R., Capelato H.V., Mazure A.
1997, ApJ 485, 447
\bibitem{} David L.P., Slyz A., Jones C., et al. 1993, ApJ 412, 479
\bibitem{} Driver S.P., Phillipps S., Davies J.I., Morgan I.,
Disney M.J. 1994, MNRAS 268, 393
\bibitem{} Durret F., Forman W., Gerbal D., Jones C., Vikhlinin A. 1998,
A\&A 335, 41
\bibitem{} Durret F., Gerbal D., Lobo C., Pichon C. 1999a, A\&A 343, 760
\bibitem{} Durret F., Felenbok P., Lobo C., Slezak E. 1999b, A\&AS 139, 525
\bibitem{} Edge A.C., Stewart G.C. 1991, MNRAS, 252, 428
\bibitem{} Fadda D., Slezak E., Bijaoui A. 1998, A\&AS 127, 335 
\bibitem{} Gurzadyan V.G., Mazure A. 1998, MNRAS 295, 177
\bibitem{} Katgert P., Mazure A., Perea J. et al. 1996, A\&A 310, 8
\bibitem{} Kriessler J.R., Beers T.C. 1997, AJ 113, 80
\bibitem{} Lima Neto G.B., Gerbal D., M\'arquez I. 1999, MNRAS 309, 481
\bibitem {} Lin H., Kirshner R.P., Schectman S.A. et al.
 1996, ApJ 464, 60
\bibitem{} Lobo C., Biviano A., Durret F., et al. 1997, A\&A 317, 385
\bibitem{} Lumsden S.L., Collins C.A., Nichol R.C., Eke V.R., Guzzo L. 1997,
MNRAS 290, 119
\bibitem{}  Markevitch M., Vikhlinin A., Forman W.R., Sarazin C.L. 1999, 
ApJ 527, 545
\bibitem{} Mazure A., Katgert P., den Hartog R. et al. 1996, A\&A 310, 31
\bibitem{}  Mohr J.J., Mathiesen B., Evrard A.E. 1999, ApJ 517, 627
\bibitem{}  Molinari E., Chincarini G., Moretti A., De Grandi S. 1998,
A\&A 338, 874
\bibitem{} Oegerle W.R., Hill J.M. 1994, AJ 107, 857
\bibitem{} Pislar V., Durret F., Gerbal D., Lima Neto G.B., Slezak E. 1997, 
A\&A 322, 53
\bibitem{} Pislar V. 1998, PhD Thesis, Universit\'e Paris 6
\bibitem{} Rauzy S., Adami C., Mazure A. 1998, A\&A 337, 31
\bibitem{} Serna A., Gerbal D. 1996, A\&A 309, 65
\bibitem{} Slezak E., Durret F., Guibert J., Lobo C. 1999, A\&AS 139, 559
\bibitem{} Snowden S.L. 1995, {\sl Cookbook for analysis procedures
for Rosat XRT/PSPC observations of extended objects and diffuse background}
(Greenbelt: NASA USRSDC)
\bibitem{} Snowden S.L., McCammon D., Burrows D.N., Mendenhall J.A. 1994, ApJ
424, 714
\bibitem{} Struble M.F., Rood H.J. 1987, ApJS 63, 543
\bibitem{} Tyson J.A., 1988, AJ 96, 1
\bibitem{} Valotto C.A., Nicotra M.A., Muriel H., Lambas D.G. 1997, ApJ 479,
90
\bibitem{} Wu X.-P., Xue Y.-J., Fang L.-Z. 1999, ApJ 524, 22
\bibitem{} Zabludoff A.I., Geller M.J., Huchra J.P., Vogeley M.S. 1993, 
AJ 106, 1273

\end{thebibliography}
\end{document}